\documentclass[aps,pra,twocolumn,superscriptaddress,reprint,a4paper,longbibliography,floatfix]{revtex4-2}

\usepackage[utf8]{inputenc}
\usepackage[T1]{fontenc}
\usepackage{amsmath,amssymb,amsfonts,amsthm}
\usepackage{mathtools}
\usepackage{graphicx}% Include figure files
\usepackage{xcolor}

\usepackage{dcolumn}% Align table columns on decimal point
\usepackage{bm}% bold math
\usepackage{dsfont}
\usepackage{tikz} %for ORCID
\usepackage{xparse} % for commands with multiple arguments
\usepackage{xspace} % for commands with math
\usepackage{hhline}

\usepackage{braket}
\usepackage{slashed}
\usepackage{subcaption}
\usepackage{booktabs}
\graphicspath{ {./figs/} }
\makeatletter
\renewcommand{\@makecaption}[2]{%
  \par\vspace{\abovecaptionskip}%
  \begingroup
    \footnotesize % Change to \footnotesize if your figures use \footnotesize
    \sbox\@tempboxa{#1: #2}%
    \ifdim \wd\@tempboxa >\hsize
      \leftskip\z@ \rightskip\z@ \parfillskip\z@ plus 1fil
      #1: #2\par
    \else
      \hb@xt@\hsize{\hfil\box\@tempboxa\hfil}\par
    \fi
  \endgroup
  \vspace{\belowcaptionskip}%
}
\makeatother

\definecolor{myurlcolor}{rgb}{0,0,0.7}
\definecolor{myrefcolor}{rgb}{0.8,0,0}
\usepackage{hyperref}
\hypersetup{
    unicode=true, bookmarksnumbered=false, bookmarksopen=false, breaklinks=false, pdfborder={0 0 0},
    colorlinks=true,
    linkcolor=myrefcolor,citecolor=myurlcolor,urlcolor=myurlcolor
}
\usepackage{cleveref}
\theoremstyle{plain}% default
\newtheorem{theorem}{Theorem}

\theoremstyle{definition}
\theoremstyle{remark}
\usepackage[normalem]{ulem}

\usepackage[compat=1.1.0]{tikz-feynman} % Modern standard

\newcommand{\prlparagraph}[1]{\emph{#1.---}}

\renewcommand{\vec}[1]{\bm{#1}}

\newcommand{\mrm}[1]{\mathrm{#1}}

\newcommand{\trm}[1]{\textrm{#1}}

\newcommand{\dd}{\mrm{d}}
\newcommand{\ee}{\mrm{e}}
\newcommand{\ii}{\mrm{i}}

\newcommand{\1}{\mathds{1}}
\newcommand{\TT}{\mathsf{T}}

\renewcommand{\bra}[1]{\left\langle#1\right|}
\renewcommand{\ket}[1]{\left|#1\right\rangle}
\newcommand{\ketbra}[2]{\ket{#1}\!\bra{#2}}

\newcommand{\proj}[1]{\ketbra{#1}{#1}}

\renewcommand{\Re}[1]{\mrm{Re}\!\left\{#1\right\}}

\newcommand{\Tr}{\mrm{Tr}}
\NewDocumentCommand\trace{g g}{%
  \IfNoValueTF{#1}
    {\Tr}
    {\IfNoValueTF{#2}
      {\Tr\!\left\{#1\right\}}
      {\Tr_{#1}\!\left\{#2\right\}}}%
}
\NewCommandCopy\tr\trace

\newcommand{\eref}[1]{(\ref{#1})}
\newcommand{\eqnref}[1]{Eq.~(\ref{#1})}

\newcommand{\figref}[1]{Fig.~\ref{#1}}
\newcommand{\fref}[1]{Fig.~\ref{#1}}
\newcommand{\tabref}[1]{Table~\ref{#1}}

\newcommand{\appref}[1]{App.~\ref{#1}}

\newcommand{\citeref}[1]{Ref.~\cite{#1}}

\newcommand{\thmref}[1]{Theorem~\ref{#1}}

\newcommand{\R}{\mathfrak{R}}
\newcommand{\Rsep}{\R_{\mrm{sep}}}
\newcommand{\pPs}{\ensuremath{p\text{-}\mathrm{Ps}}\xspace}

\newcommand{\myindexsize}{\scriptscriptstyle}
\newcommand{\rhoE}[2]{{\rho}_{\myindexsize #1}^{\myindexsize #2}}

\newcommand{\s}[1]{\hat{\sigma}_{#1}}
\newcommand{\sx}{\s{x}}
\newcommand{\sy}{\s{y}}
\newcommand{\sz}{\s{z}}

\makeatletter

\definecolor{lime}{HTML}{A6CE39}
\DeclareRobustCommand{\orcidicon}{
	\begin{tikzpicture}
	\draw[lime, fill=lime] (0,0) 
	circle [radius=0.16] 
	node[white] {{\fontfamily{qag}\selectfont \tiny ID}};
	\draw[white, fill=white] (-0.0625,0.095) 
	circle [radius=0.007];
	\end{tikzpicture}
	\hspace{-2mm}
}
\foreach \x in {A, ..., Z}{\expandafter\xdef\csname orcid\x\endcsname{\noexpand\href{https://orcid.org/\csname orcidauthor\x\endcsname}
			{\noexpand\orcidicon}}
}
\makeatother
\begin{document}

\title{Rigorously certifying entanglement in two-photon electron--positron annihilation}

\newcommand{\IFPAN}{Institute of Physics, Polish Academy of Sciences, Aleja Lotnik\'{o}w 32/46, 02-668 Warsaw, Poland.}

\author{Przemys\l{}aw Czapla\orcidB{}}
\email{pczapla@ifpan.edu.pl}
\affiliation{\IFPAN}

\author{Tomasz Sowi\'{n}ski\orcidC{}}
\email{tomsow@ifpan.edu.pl}
\affiliation{\IFPAN}

\author{Jan Ko\l{}ody\'{n}ski\orcidA{}}
\email{jankolo@ifpan.edu.pl}
\affiliation{\IFPAN}

\date{\today}

\begin{abstract}
By constructing the quantum measurement realised by Compton scattering of an arbitrary back-to-back two-photon state from electron--positron annihilation, we rigorously certify its entanglement. Going beyond para-positronium decay, whose QED symmetries render the problem trivial, we identify a hierarchy of line-of-flight rotational symmetries and conditions under which the commonly used $\R$-factor is the optimal entanglement witness. This allows us to reinterpret recent J-PET measurements of $\R$ and lower-bound the negativity of the two-photon states experimentally produced.
\end{abstract}

\maketitle

%%%%%%%%%%%%%%%%%%%%%%%%%%%%%%%%%%%%%%%%%%%%%%%%%%%%%%%%%%%%%%%%%%%%%%%%%%%%%%%%%%%%%%%%%%%
%%%%%%%%%%%%%%%%%%%%%%%%%%%%%%%%%%%%%%%%%%%%%%%%%%%%%%%%%%%%%%%%%%%%%%%%%%%%%%%%%%%%%%%%%%%
\prlparagraph{Introduction}
Two-photon annihilation of electron--positron pairs at rest produces back-to-back photons, whose polarisations are predicted by quantum electrodynamics (QED) to be maximally entangled for para-positronium (\pPs) decay~\cite{Wheeler1946,Pryce1947}. Since the pioneering experiments~\cite{Snyder1948,Bleuler1948,Hanna1948,Wu1950a}, the photons' polarisation correlations have been probed via Compton scattering and quantified by the $\R$-factor, recently revisited in the context of decoherence~\cite{Strizhak2022,Bordes2024,Parashari2024}, positron emission tomography~\cite{Watts2021,Moskal2025} and entanglement witnessing~\cite{Hiesmayr2019,Caradonna2019,Joos2026}.

Yet, the role of entanglement has so far been assessed either by comparing a measured $\R$ with values predicted for a few selected separable states~\cite{Snyder1948,Bohm1957,Peres1960}, or by inspecting the consistency of the scattering distribution with the maximally entangled state~\cite{Joos2026}. Neither establishes a bound on $\R$ valid for \emph{all} separable states, as required to rigorously certify entanglement, i.e.~relying only on the measurement model, as in quantum information~\cite{Horodecki2009,Guehne2009} and high-energy physics~\cite{ashby2023,Afik2025}.

Here, starting from QED, we derive the Compton-scattering cross-section for an arbitrary back-to-back two-photon state and show that it defines a quantum measurement, i.e.~a positive operator-valued measure (POVM), on the photons' polarisations. We prove that $\R$ is a valid entanglement witness for any such state, with a tight separability threshold that coincides with the historical ``classical'' benchmarks. Identifying a hierarchy of rotational symmetries about the line of flight, we further show that for locally azimuthally symmetric states $\R$ is optimal whenever the coherence it probes dominates, i.e.~no other analysis of Compton-scattering data can then do better. Finally, as $\R$ lower-bounds the negativity, we quantify the entanglement in recent J-PET data~\cite{Moskal2025} without any assumption on the state, its source, or the polar-angle dependence of the detection efficiency.

\begin{figure}
    \centering
    \includegraphics[width=\columnwidth]{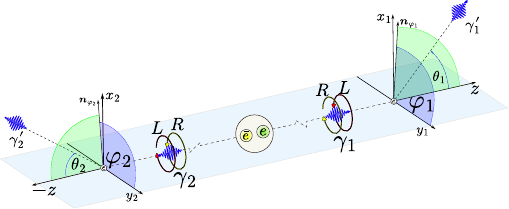}
    \caption{\textbf{Compton scattering of annihilation photons}. Photons $\gamma_{1,2}$, emitted back-to-back along $\pm z$ upon two-photon electron--positron annihilation at rest (e.g.~\pPs decay), Compton-scatter into $\gamma'_{1,2}$. Polar angles $\theta_{1,2}$ lie within the scattering planes (green), oriented relative to $y_{1,2}$ by azimuthal angles $\varphi_{1,2}$ (blue).}
    \label{fig:scattering_geometry}
\end{figure}

%%%%%%%%%%%%%%%%%%%%%%%%%%%%%%%%%%%%%%%%%%%%%%%%%%%%%%%%%%%%%%%%%%%%%%%%%%%%%%%%%%%%%%%%%%%
%%%%%%%%%%%%%%%%%%%%%%%%%%%%%%%%%%%%%%%%%%%%%%%%%%%%%%%%%%%%%%%%%%%%%%%%%%%%%%%%%%%%%%%%%%%
\prlparagraph{Annihilation photons as a two-qubit system}%
When an electron and a positron annihilate at rest into two photons, energy and linear momentum conservation in the centre-of-mass frame dictate that their respective momenta satisfy $\vec{k}_1 = \vec{k}$ and $\vec{k}_2 = -\vec{k}$, i.e.~the photons propagate back-to-back along $+z$ and $-z$, respectively, with the $z$-axis chosen along the line of flight (see \figref{fig:scattering_geometry}), each carrying energy equal to the electron rest energy, $E =\hbar c |\vec{k}| = m_e c^2 \approx 511\,\mathrm{keV}$.

The helicity basis naturally describes the photons' polarisation degrees of freedom. Specifically, the polarisation vector $\vec{\epsilon}_\lambda(\vec{k})$ for a photon propagating with momentum $\vec{k}$ and helicity $\lambda = \pm 1$ (corresponding to right-handed $\ket{R}$ and left-handed $\ket{L}$ circular-polarisation states~%
\footnote{As in particle physics~\cite{Peskin:1995} and engineering~\cite{IEEE145}, but contrary to classical optics~\cite{BornWolf,Jackson}.}%
) is identified, up to a phase, by $\vec{\epsilon}_\lambda(\vec{k}) \times \vec{n}_{\vec{k}} = \ii \lambda\, \vec{\epsilon}_\lambda(\vec{k})$ with $\vec{n}_{\vec{k}}\equiv\vec{k}/|\vec{k}|$. Each choice of $(\vec{k}, \lambda)$ completely characterises a single plane-wave mode of the electromagnetic field. Thus, in second quantisation, bosonic annihilation (creation) operators $\hat{a}_{\vec{k},\lambda}$ ($\hat{a}_{\vec{k},\lambda}^\dagger$) can be defined for each such mode, satisfying $[\hat{a}_{\vec{k},\lambda},\hat{a}_{\vec{k}',\lambda'}^\dagger]=\delta^{(3)}(\vec{k}-\vec{k}')\,\delta_{\lambda\lambda'}$~\cite{Bialynicki2017}. Consequently, for fixed momenta $\pm\vec{k}$, the two-photon state space can be identified with the standard two-qubit Hilbert space $\mathbb{C}^2\otimes\mathbb{C}^2$, in which the entanglement can be examined~\cite{Horodecki2009}. We fix the orthonormal basis of this space, which we refer to as the \emph{circular (polarisation) basis}, to be, in order:
\begin{subequations}
\label{eq:RL_basis}
  \begin{align}
      \hat{a}^{\dagger}_{\vec{k},+1}\hat{a}^{\dagger}_{-\vec{k},+1} \ket{0}&=\ket{R}_1\ket{R}_2 \equiv \ket{RR},\\ %\;\eqqcolon \dket{1},\\
      \hat{a}^{\dagger}_{\vec{k},+1}\hat{a}^{\dagger}_{-\vec{k},-1} \ket{0}&=\ket{R}_1\ket{L}_2 \equiv \ket{RL},\\ %\;\eqqcolon \dket{2},\\
      \hat{a}^{\dagger}_{\vec{k},-1}\hat{a}^{\dagger}_{-\vec{k},+1} \ket{0}&=\ket{L}_1\ket{R}_2 \equiv \ket{LR},\\ %\;\eqqcolon \dket{3},\\
      \hat{a}^{\dagger}_{\vec{k},-1}\hat{a}^{\dagger}_{-\vec{k},-1} \ket{0}&=\ket{L}_1\ket{L}_2 \equiv \ket{LL}, %\;\eqqcolon \dket{4},
  \end{align}
\end{subequations}
and parametrise the most general two-photon density matrix, $\rho\in\mathcal{B}(\mathbb{C}^2\otimes\mathbb{C}^2)$ with $\rho\ge0$ and $\Tr{\rho}=1$, as~\cite{Bengtsson2006}
\begin{equation}
  \rho 
    = \!\!\!\!\! \sum_{\substack{{\scriptscriptstyle i,j \in \{R,L\}} \\ {\scriptscriptstyle i',j' \in \{R,L\}}}} \!\!\!\!\! \rho_{i'j'}^{ij}\ketbra{ij}{i'j'} 
    =
   \begin{pmatrix}
      \rhoE{RR}{RR} & \rhoE{RL}{RR} & \rhoE{LR}{RR} & \rhoE{LL}{RR} \\[0.2em]
      \rhoE{RR}{RL} & \rhoE{RL}{RL} & \rhoE{LR}{RL} & \rhoE{LL}{RL} \\[0.2em]
      \rhoE{RR}{LR} & \rhoE{RL}{LR} & \rhoE{LR}{LR} & \rhoE{LL}{LR} \\[0.2em]
      \rhoE{RR}{LL} & \rhoE{RL}{LL} & \rhoE{LR}{LL} & \rhoE{LL}{LL}
    \end{pmatrix}.
    \label{eq:gen_mixed}
\end{equation}

%%%%%%%%%%%%%%%%%%%%%%%%%%%%%%%%%%%%%%%%%%%%%%%%%%%%%%%%%%%%%%%%%%%%%%%%%%%%%%%%%%%%%%%%%%%
%%%%%%%%%%%%%%%%%%%%%%%%%%%%%%%%%%%%%%%%%%%%%%%%%%%%%%%%%%%%%%%%%%%%%%%%%%%%%%%%%%%%%%%%%%%
\prlparagraph{QED vs rotational symmetries}%
If the two-photon polarisation state~\eref{eq:gen_mixed} results from \pPs decay, the QED symmetries fully determine its form~\cite{Yang1950}. With \pPs in its ground state (${}^1S_0$), the electron ($e$) and positron ($\bar{e}$) form a spin singlet, $\frac{1}{\sqrt{2}}\left(\ket{\uparrow}_{e}\ket{\downarrow}_{\bar{e}} - \ket{\downarrow}_{e}\ket{\uparrow}_{\bar{e}}\right)$, with orbital angular momentum $L=0$ and spin $S=0$, locking its total angular momentum to $J=0$. Under \emph{charge conjugation} $\mathcal{C}$, a fermion-antifermion bound state carries the eigenvalue $C = (-1)^{L+S} = +1$. Since an $n$-photon system has $\mathcal{C}$-parity $(-1)^n$, $\mathcal{C}$-conservation forbids decay into an odd number of photons, making the two-photon channel ($n=2$) dominant.

Although total angular momentum conservation severely restricts the allowed two-photon state, it is \emph{spatial parity} ($\mathcal{P}$) conservation that singles out a unique one. With the quantisation axis along $z$, the vanishing projection $J_z = 0$ confines the photons to identical local circular polarisations (equal helicities), i.e.~to $\trm{span}\{\ket{RR}, \ket{LL}\}$ (see \eqnref{eq:RL_basis}), with full rotational invariance ($J=0$) imposing no further constraint~\cite{Yang1950,JacobWick1959}. Parity, with the phases fixed by $\vec{\epsilon}_\lambda(-\vec{k})=\vec{\epsilon}_{-\lambda}(\vec{k})$~\cite{JacobWick1959}, acts as $\mathcal{P}\hat{a}^\dagger_{\vec{k},\lambda}\mathcal{P}^{-1}=-\hat{a}^\dagger_{-\vec{k},-\lambda}$, so that $\mathcal{P}\ket{RR}=\ket{LL}$, $\mathcal{P}\ket{RL}=\ket{RL}$ and $\mathcal{P}\ket{LR}=\ket{LR}$. Since \pPs has $P=(-1)^{L+1}=-1$, the only allowed state is the odd-parity Bell state~\cite{Nielsen2000}, referred to as the singlet state in this context~\cite{Snyder1948}:
\begin{equation}
    \ket{\Phi_-} = \tfrac{1}{\sqrt{2}} \left( \ket{R}_1 \ket{R}_2 - \ket{L}_1 \ket{L}_2 \right).
    \label{eq:Phi_-}
\end{equation}

Note that the state \eref{eq:Phi_-} in the linear-polarisation basis~\cite{Bohm1957,Bohm_1960}, defined by $\ket{H}=\tfrac{1}{\sqrt{2}}(\ket{R}+\ket{L})$ and $\ket{V}=\tfrac{\ii}{\sqrt{2}}(\ket{R}-\ket{L})$~\cite{James2001}, reads, up to a global phase, $\tfrac{1}{\sqrt{2}}\left(\ket{H}_1\ket{V}_2+\ket{V}_1\ket{H}_2\right)$~\cite{Caradonna2019}---a form reminiscent of the two-photon state generated via type-II spontaneous parametric down-conversion~\cite{Kwiat1995}. As $\ket{\Phi_-}$ is maximally entangled~\cite{Horodecki2009}, assuming \pPs decay as the only photon source and imposing QED symmetries renders any further entanglement analysis pointless~\cite{Peres1960}, with realistic, noisy experimental scenarios~\cite{Strizhak2022,Bordes2024,Parashari2024} requiring a broader class of two-photon states to be considered.

Importantly, the entanglement certification methods we introduce below apply to any two-photon polarisation state~\eref{eq:gen_mixed}, regardless of its origin, provided that each event consists of exactly two photons emitted back-to-back~%
\footnote{For \pPs, higher-order even-photon channels ($n=4$) are suppressed by $\mathcal{O}(\alpha^2)$.}. Within this general setting, we identify geometrically motivated \emph{rotational symmetries} about the line of flight, which define a hierarchy of three nested classes of states, $\mathrm{LAS}\supset C_4\supset\mathrm{GAS}$:

\noindent
\textbf{(LAS)}~%
\emph{Locally azimuthally symmetric} states, for which the reduced state $\rho^{(\ell)}\coloneqq\tr{\bar\ell}{\rho}$ of each photon, $\ell\in\{1,2\}$ with $\bar\ell\neq\ell$, is invariant under rotations about $\pm z$ and, hence, does not carry any linear polarisation, i.e.~$\bra{R}\rho^{(\ell)}\ket{L}=0$~%
\footnote{In the linear-polarisation basis, $\bra{R}\rho^{(\ell)}\ket{L}=\tfrac12\big(S^{(\ell)}_1+\ii S^{(\ell)}_2\big)$, with $S^{(\ell)}_1\coloneqq\bra{H}\rho^{(\ell)}\ket{H}-\bra{V}\rho^{(\ell)}\ket{V}$ and $S^{(\ell)}_2\coloneqq2\Re{\bra{H}\rho^{(\ell)}\ket{V}}$ being the linear Stokes parameters~\cite{James2001}.}. In terms of \eqnref{eq:gen_mixed}, these conditions read $\rhoE{LR}{RR}+\rhoE{LL}{RL}=0$ ($\ell=1$) and $\rhoE{RL}{RR}+\rhoE{LL}{LR}=0$ ($\ell=2$), and can be verified experimentally via flat single-photon azimuthal scattering distributions.

\noindent
\textbf{(\boldmath$C_4$)}~%
\emph{$C_4$-symmetric} states, which are invariant under joint rotations of both photons by $\pi/2$ about the line of flight---a symmetry that may persist when the global azimuthal symmetry is broken, e.g.~by a preferred linear-polarisation basis~\cite{Furry1936,Peres1960} or by a segmented detector geometry. As $\ket{RR}$ and $\ket{LL}$ span the $J_z=0$ sector, while $\ket{RL}$ and $\ket{LR}$ carry $J_z=\pm2$, such rotations forbid all coherences between sectors whose $J_z$ differ by $\pm2$, so that $\rhoE{LR}{RR}=\rhoE{LL}{RL}=\rhoE{RL}{RR}=\rhoE{LL}{LR}=0$, i.e.~they are so-called \emph{X-states}~\cite{Yu2005} in the circular basis \eref{eq:RL_basis}.

\noindent
\textbf{(GAS)}~%
\emph{Globally azimuthally symmetric} states, which are invariant under joint rotations of both photons by any angle about the line of flight ($J_z$ conservation), form the subclass of X-states that additionally satisfy $\rhoE{LR}{RL}=0$~%
\footnote{If $C_n$ symmetry with $n\neq4$ is imposed instead, $C_2$ leaves the general state \eref{eq:gen_mixed} unconstrained, while $C_3$ and $C_{n>4}$ already enforce $\rhoE{LR}{RL}=0$, as for GAS states.}.

%%%%%%%%%%%%%%%%%%%%%%%%%%%%%%%%%%%%%%%%%%%%%%%%%%%%%%%%%%%%%%%%%%%%%%%%%%%%%%%%%%%%%%%%%%%
%%%%%%%%%%%%%%%%%%%%%%%%%%%%%%%%%%%%%%%%%%%%%%%%%%%%%%%%%%%%%%%%%%%%%%%%%%%%%%%%%%%%%%%%%%%
\prlparagraph{Compton scattering as a POVM}%
Given a pure two-photon input state $\rho = \proj{\psi}$ in \eqnref{eq:gen_mixed}, the joint probability of detecting both photons at solid angles $\Omega_\ell=(\theta_\ell,\varphi_\ell)$ (see \figref{fig:scattering_geometry}) after undergoing Compton scattering off stationary electrons ($\vec{k}_{e_\ell}\!=\!0$) in initial states $\ket{\phi_0}_{e_\ell}$ is determined by the squared modulus of the (total) transition amplitude, i.e.~the $\hat{S}$-matrix element $\bra{\Psi_\mrm{out}}\hat{S}\ket{\Psi_\mrm{in}}$~\cite{Peskin:1995}:
\begin{align}
  & P\left(\ket{\xi}\!, K_1, K_2 \Bigm| \ket{\psi}, k^\mu; \ket{\phi_0}, s_1, s_2 \right) = \label{eq:scattering_prob} \\
  & \; \Big| \underbrace{\bigotimes_{\ell=1}^2 \,{}_{\gamma_\ell,e_\ell}\!\bra{\xi(K_\ell)}}_{\bra{\Psi_\mrm{out}}}
  \raisebox{-1.5pt}{\scalebox{1.5}{$\hat{S}$}}
  \underbrace{\ket{\psi(k^\mu)}_{\gamma_1,\gamma_2} \bigotimes_{\ell=1}^2 \ket{\phi_0(\vec{k}_{e_\ell}\!=\!0, s_\ell)}_{e_\ell}}_{\ket{\Psi_\mrm{in}}} \Big|^2\!, \nonumber
\end{align}
where by $k^{\mu}$ ($k'^{\mu}$) we denote the momentum four-vectors of incoming (outgoing) particles and by $s_\ell$ ($s'_\ell$) the electron spin labels. $\ket{\xi(K_\ell)}_{\gamma_\ell,e_\ell}$ is a general joint state of photon $\gamma_\ell$ and electron $e_\ell$ following the scattering event, and is parametrised by $K_\ell \coloneqq \{k'^{\mu}_{\gamma_\ell}(\Omega_\ell); k'^\mu_{e_\ell}, s'_\ell\}$ with the photon being scattered into the solid angle $\Omega_\ell$.

% To obtain the desired distribution over solid angles $\vec{\Omega}=\{\Omega_1,\Omega_2\}$, \eqnref{eq:scattering_prob} must be integrated over the unobserved output spatial momenta---subject to energy-momentum conservation---and traced over both initial and final electron spin degrees of freedom, with the former assumed to be unpolarised (maximally mixed), i.e.~$p(s_1,s_2)=1/4$~\cite{Peskin:1995}. Integrating over the output momenta and averaging over electron spin values, $\sum_{s_1,s_2,s_1',s_2'} p(s_1,s_2) \int\! \dd^3\vec{k}_{\gamma_1}' \dd^3\vec{k}_{\gamma_2}' \dd^3\vec{k}_{e_1}' \dd^3\vec{k}_{e_2}' \delta^{(4)}(k'^{\mu}_{\gamma_1}+k'^\mu_{e_1}-k^\mu) \,\delta^{(4)}(k'^{\mu}_{\gamma_2}+k'^\mu_{e_2}-k^\mu) \dots$, we recast the detection probability \eref{eq:scattering_prob} into the standard Born-rule trace form:

To obtain the distribution over solid angles $\vec{\Omega}=\{\Omega_1,\Omega_2\}$, \eqnref{eq:scattering_prob} must be integrated over the unobserved momenta, i.e.~the magnitudes of the scattered-photon momenta and the recoil-electron momenta, subject to energy-momentum conservation, and traced over the initial and final electron spin degrees of freedom, with the former unpolarised (maximally mixed), i.e.~$p(s_1,s_2)=1/4$~\cite{Peskin:1995}. This recasts the detection probability \eref{eq:scattering_prob} into the standard Born-rule trace form:
\begin{equation}
    P(\vec{\Omega} \mid \rho^{|\vec{k}|}) = \tr{\rho^{|\vec{k}|}\,M_{\vec{\Omega}}^{|\vec{k}|}}.
    \label{eq:POVM_trace}
\end{equation}
This allows us to extend the formalism to any mixed two-photon polarisation state $\rho^{|\vec{k}|}=\sum_i\lambda_i\proj{\psi_i^{\vec{k}^{(i)}}}$ in \eqnref{eq:gen_mixed}, even when data are aggregated over events with different directions $\vec{k}^{(i)}$, as all of them share the same $|\vec{k}|$. Moreover, we identify
\begin{equation}
    M_{\vec{\Omega}}^{|\vec{k}|}\coloneqq \frac{1}{4} \sum_{s,s'} {E^{|\vec{k}|}_{\vec{\Omega}}(s,s')}^\dagger E^{|\vec{k}|}_{\vec{\Omega}}(s,s')
    \label{eq:POVM_element}
\end{equation}
as the elements of a positive operator-valued measure (POVM) satisfying $M_{\vec{\Omega}}^{|\vec{k}|}\ge0$ for all $\vec{\Omega}$ and, conditioned on both photons being Compton-scattered (see \appref{app:witnesses}), $\int\!\dd\vec{\Omega} \, M_{\vec{\Omega}}^{|\vec{k}|} \!=\! \1$, with measurement operators given by
\begin{equation}
    E^{|\vec{k}|}_{\vec{\Omega}}(s,s') = \bigotimes_{\ell=1}^2 \,{}_{\gamma_\ell,e_\ell}\!\bra{\xi(\Omega_\ell,s'_\ell)} 
    \;\raisebox{-1.5pt}{\scalebox{1.5}{$\hat{S}$}}\;
    \bigotimes_{\ell=1}^2\ket{\phi_0(s_\ell)}_{e_\ell}.
    \label{eq:measurement_ops}
\end{equation}

Crucially, energy-momentum conservation in two-photon annihilation at rest locks the photon momentum magnitude to $|\vec{k}|=m_e c/\hbar$ across every event. This guarantees that all scattering events share an identical two-qubit Hilbert space $\mathbb{C}^2\otimes\mathbb{C}^2$, and static POVM elements $M_{\vec{\Omega}}^{|\vec{k}|}$, in sharp contrast to channels producing more photons (such as ortho-positronium, $o\text{-Ps}$) where continuous kinetic fluctuations make event-by-event entanglement certification significantly more challenging~\cite{Acin2001,Hiesmayr2017,Dulski2021,Moskal2021}.

%%%%%%%%%%%%%%%%%%%%%%%%%%%%%%%%%%%%%%%%%%%%%%%%%%%%%%%%%%%%%%%%%%%%%%%%%%%%%%%%%%%%%%%%%%%
%%%%%%%%%%%%%%%%%%%%%%%%%%%%%%%%%%%%%%%%%%%%%%%%%%%%%%%%%%%%%%%%%%%%%%%%%%%%%%%%%%%%%%%%%%%
\prlparagraph{Differential cross-section and the $\R$-factor}
Scattering theory in QED allows one to directly evaluate the unnormalised counterpart of \eqnref{eq:POVM_trace} in the form of the differential cross-section, $\dd\sigma$~\cite{Peskin:1995}. By decomposing the $S$-matrix to isolate the scattering part, $\hat{S} = \hat{\1} + \ii\hat{T}$, and assuming that the two photons scatter off distinct electrons, the joint $\hat{T}$-matrix factorises as $\hat{T}=\bigotimes_{\ell=1}^2\hat{T}_\ell$. For each photon, we derive in \appref{app:scat_deriv} the spin-averaged effective scattering operator, whose elements expressed in the circular basis ($i,j\in\{R,L\}$) read $[\hat{\Sigma}_\ell(\Omega_\ell)]^j_{i}\coloneqq\frac{1}{2}\sum_{s',s} \bra{\Omega_\ell;s'}\hat{T}_\ell\ket{i;s} \bra{j;s}\hat{T}_\ell^\dagger\ket{\Omega_\ell;s'}$ (up to a kinematic prefactor), where $s$ ($s'$) label the initial (final) electron spin states, and the final scattered photon state $\ket{\Omega_\ell}$ is determined by the solid angle $\Omega_\ell$---recall \figref{fig:scattering_geometry}. Consequently, the differential cross-section for a general two-photon mixed state \eref{eq:gen_mixed} takes the form (see \appref{app:scat_deriv}):
\begin{align}
    \frac{\dd^2\sigma(\rho)}{\dd\Omega_1\dd\Omega_2} 
    = \tr{\rho\,\hat{\Sigma}_1(\Omega_1)\otimes\hat{\Sigma}_2(\Omega_2)^*}\!, 
    \label{eq:psi_cross} 
\end{align}
where the effective scattering operator acting on the (input) polarisation space of each photon reads
\begin{equation}
    \label{eq:Sigma_l}
    \hat{\Sigma}_\ell(\Omega_\ell) =  U_z(\varphi_\ell)
        \begin{pmatrix}
            A(\theta_\ell) & B(\theta_\ell) \\
            B(\theta_\ell) & A(\theta_\ell)
        \end{pmatrix}    
        U_z^\dagger(\varphi_\ell),
\end{equation}
with $U_z(\varphi)\coloneqq\ee^{\ii \varphi \sz}$ being the rotation unitary matrix in the circular basis, and
\begin{equation}
       A(\theta) \coloneqq \frac{r_e^2}{4}\,\varepsilon^2(\varepsilon+\varepsilon^{-1}-\sin^2\theta), 
       \quad 
       B(\theta) \coloneqq -\frac{r_e^2}{4}\,\varepsilon^2\sin^2\theta,
       \label{eq:AB}
\end{equation}
where $\varepsilon\coloneqq k'^0/k^0=1/(2-\cos\theta)$ and $r_e=e^2/(4\pi\varepsilon_0 m_e c^2)$ is the classical electron radius.

Substituting the parametrisation of $\rho$ introduced in \eqnref{eq:gen_mixed} into \eqnref{eq:psi_cross}, we obtain
\begin{align}
    \frac{\dd^2\sigma(\rho)}{\dd\Omega_1\dd\Omega_2} 
    & = X + 2Y \big[ |\rhoE{LL}{RR}| \cos(2(\varphi_1-\varphi_2) - \arg\{\rhoE{LL}{RR}\}) \nonumber \\
    & \qquad\qquad\quad + |\rhoE{LR}{RL}| \cos(2(\varphi_1+\varphi_2) - \arg\{\rhoE{LR}{RL}\}) \big] \nonumber\\
    & \quad + 2 A(\theta_2) B(\theta_1)\, \Re{\bra{R}\rho^{(1)}\ket{L}\, \ee^{-2\ii\varphi_1}} \nonumber\\
    & \quad + 2 A(\theta_1) B(\theta_2)\, \Re{\bra{R}\rho^{(2)}\ket{L}\, \ee^{2\ii\varphi_2}},
    \label{eq:gen_cross_section}
\end{align}
where $X\coloneqq A(\theta_1) A(\theta_2)$ and $Y\coloneqq B(\theta_1) B(\theta_2)$. The last two terms in \eqnref{eq:gen_cross_section} are local---they stem from the linear polarisation of the individual photons and vanish identically if and only if the state is LAS.

However, in typical experimental scenarios, only the relative azimuthal angle $\Delta\varphi\coloneqq\varphi_1-\varphi_2$ is recorded for each scattering event. To account for this, we substitute $\varphi_1\to\Delta\varphi+\varphi_2$ in \eqnref{eq:gen_cross_section} and average the differential cross-section over the unobserved angle $\varphi_2$, upon which both the local terms and the term involving $\rhoE{LR}{RL}$ vanish:
\begin{align}
    \left\langle\frac{\dd^2\sigma(\rho)}{\dd\Omega_1\dd\Omega_2}\right\rangle 
    & = X + 2Y |\rhoE{LL}{RR}| \cos(2\Delta\varphi-\arg\{\rhoE{LL}{RR}\}).
    \label{eq:av_cross_section}
\end{align}

Fitting the experimental data to the $\Delta\varphi$-modulation in \eqnref{eq:av_cross_section} and taking its maximum-to-minimum ratio yields the so-called \emph{$\R$-factor}:
\begin{equation}
  \R(\rho) \coloneqq  \frac{X+2Y|\rhoE{LL}{RR}|}{X-2Y|\rhoE{LL}{RR}|},
  \label{eq:R_factor}
\end{equation}
where the parameters $X$ and $Y$ are determined by the measurement, i.e.~the polar angles $\theta_\ell$. Introduced by \citet{Snyder1948}, $\R$ has served as the key benchmark in landmark~\cite{Bleuler1948, Hanna1948, Wu1950a, Langhoff1960, Kasday1975, Wilson1976} and recent~\cite{Watts2021, Strizhak2022, Bordes2024, Parashari2024, Moskal2025} experiments.

Historically, two-photon decay of \pPs was understood to produce the singlet state \eref{eq:Phi_-}, for which the differential cross-section and the resulting ratio $\R$ were derived already in the foundational works~\cite{Wheeler1946, Pryce1947, Snyder1948}. These predictions were compared with the product state $\ket{H}_1\ket{V}_2$, proposed as a ``classical'' benchmark~\cite{Snyder1948}, despite its violation of the requisite $J$ (incl.~$J_z$) and $\mathcal{P}$ symmetries. \citet{Bohm1957} instead postulated that azimuthal ($J_z$) symmetry should at least be preserved, proposing either the product state $\ket{R}_1\ket{R}_2$ or the mixed state obtained by azimuthally averaging $\ket{H}_1\ket{V}_2$ as valid reference states. In parallel, motivated by Furry's decoherence hypothesis~\cite{Furry1936}---which arbitrarily singles out the linear-polarisation basis as preferred~\cite{Caradonna2019, Caradonna2024}---the ($C_4$-symmetric) equal mixture of $\ket{H}_1\ket{V}_2$ and $\ket{V}_1\ket{H}_2$ was proposed as an alternative benchmark~\cite{Peres1960}. As shown in \tabref{tab:states_R_examples}, substituting the corresponding value of $|\rhoE{LL}{RR}|$ for each of these states into \eqnref{eq:R_factor} consistently recovers the $\R$-factor values previously established in the literature.

As we show below, access to $|\rhoE{LL}{RR}|$ via \eqnref{eq:R_factor} \emph{suffices to rigorously certify entanglement} of a general state~\eref{eq:gen_mixed}. Contrary to previous claims~\cite{Hiesmayr2019}, local scattering cross-sections can be used to perform partial or even full tomography of the quantum state and to verify correlations between the decay products~\cite{ashby2023,Aguilar2024,Afik2025,Cheng2026}. This is because they effectively implement local measurements whose elements form an over-complete set of operators, i.e.~not restricted to a single measurement basis~\cite{Guehne2002,Guehne2003}.

\begin{table}[t!]
    \begin{ruledtabular}
        \begin{tabular}{lcc}
        \rule[-1.5ex]{0pt}{4.2ex}State & $|\rhoE{LL}{RR}|$ & $\R$ \\
        \hline
        \rule{0pt}{3.2ex}Singlet \eqref{eq:Phi_-}: $\ket{\Phi_-}$ & $\tfrac{1}{2}$ & $\frac{X+Y}{X-Y}$ \\[2ex]
        \rule{0pt}{2ex}Product: $\ket{RR}\equiv\ket{R}_1\ket{R}_2$  & $0$ & $1$ \\[2ex]
        $\left.\begin{tabular}{@{}l@{}}
          Product: $\ket{HV}\equiv\ket{H}_1\ket{V}_2$ \\[1.2ex]
          $C_4$-symmetric mixture: \\
          $\tfrac{1}{2}\left(\ket{HV}\bra{HV}+\ket{VH}\bra{VH}\right)$ \\[1.2ex]
          Azimuthally averaged product: \\
          $\int_0^{2\pi}\!\!\tfrac{\dd\alpha}{2\pi}\ket{HV}_\alpha\!\bra{HV}$
        \end{tabular}\right\}$ & $\displaystyle\frac{1}{4}$ & $\displaystyle\frac{X+\tfrac{Y}{2}}{X-\tfrac{Y}{2}}$ \\[2.5ex]
        \end{tabular}
    \end{ruledtabular}
    \caption{Values of the $\R$-factor \eref{eq:R_factor} and $|\rhoE{LL}{RR}|$ in \eqnref{eq:gen_mixed} for the singlet state \eref{eq:Phi_-} and separable states considered previously in the literature as ``classical'' benchmarks~\cite{Wheeler1946, Pryce1947, Snyder1948, Bohm1957, Peres1960}. All states listed, except for $\ket{HV}$, are LAS: $\ket{\Phi_-}$, $\ket{RR}$ and the azimuthally averaged product are GAS, while the equal mixture of $\ket{HV}$ and $\ket{VH}$ is $C_4$-symmetric but not GAS.}
\label{tab:states_R_examples}
\end{table}

%%%%%%%%%%%%%%%%%%%%%%%%%%%%%%%%%%%%%%%%%%%%%%%%%%%%%%%%%%%%%%%%%%%%%%%%%%%%%%%%%%%%%%%%%%%
%%%%%%%%%%%%%%%%%%%%%%%%%%%%%%%%%%%%%%%%%%%%%%%%%%%%%%%%%%%%%%%%%%%%%%%%%%%%%%%%%%%%%%%%%%%
\prlparagraph{Detecting entanglement with the help of $\R$}
We first prove that $\R$ serves as an entanglement witness for \emph{any} state~\eref{eq:gen_mixed}, and then turn to experimentally relevant examples within the LAS class.
\begin{theorem}[All two-qubit states]
\label{thm:max_r}
For any two-photon state $\rho$ in \eqnref{eq:gen_mixed}, the coherence $\rhoE{LL}{RR}=\bra{RR}\rho\ket{LL}$ satisfies $|\rhoE{LL}{RR}|\le 1/2$. Consequently, for any scattering geometry the $\R$-factor \eref{eq:R_factor} obeys
\begin{equation}
    \R \le \R_{\max} \coloneqq \frac{X + Y}{X - Y},
    \label{eq:t1_max_r}
\end{equation}
with equality if and only if $\rho=\proj{\psi_\xi}$ is a maximally entangled state $\ket{\psi_\xi} = \frac{1}{\sqrt{2}}(\ket{RR} + \ee^{\ii\xi}\ket{LL})$ with $\xi\in\mathbb{R}$.
\end{theorem}
\begin{proof}
Positivity and unit trace of any two-qubit $\rho$ imply:
\begin{equation}
    |\rhoE{LL}{RR}| \le \sqrt{\rhoE{RR}{RR}\rhoE{LL}{LL}} \le \frac{\rhoE{RR}{RR}+\rhoE{LL}{LL}}{2} \leq \frac{1}{2}.
\end{equation}
Equality throughout requires $\rhoE{RR}{RR}=\rhoE{LL}{LL}=|\rhoE{LL}{RR}|=1/2$. Positivity then forces all matrix elements outside the $\{\ket{RR},\ket{LL}\}$ block to vanish, and this block has rank one. Hence, the bound is saturated precisely by the pure maximally entangled states $\ket{\psi_\xi}$. Substituting $|\rhoE{LL}{RR}|=1/2$ into \eqnref{eq:R_factor} yields \eqnref{eq:t1_max_r}.
\end{proof}

While the value $\R_{\max}$ was derived for the singlet state already in the foundational works~\cite{Wheeler1946, Pryce1947, Snyder1948}, \thmref{thm:max_r} proves that it is the largest possible over all two-photon states for a given scattering geometry and is attained only by maximally entangled states (cf.~\citeref{Joos2026}). As the states $\ket{\psi_\xi}$ are LAS, attaining $\R_{\max}$ also excludes any single-photon azimuthal modulation in \eqnref{eq:gen_cross_section}; cf.~\citeref{Zugec2026}, where both are nonetheless combined artificially for the purpose of semi-classical simulations. 

Moreover, as the subset of separable states is compact~\cite{Bengtsson2006} and $\R$ is continuous, $\R$ attains a maximum over separable states, which by \thmref{thm:max_r} must lie strictly below $\R_{\max}$. Hence, there exists a threshold $\R$-value above which entanglement is assured:
\begin{theorem}[Separable two-qubit states]
\label{thm:sep_r}
    Every separable two-qubit state $\rho$ in \eqnref{eq:gen_mixed} satisfies $|\rhoE{LL}{RR}|\leq 1/4$. Consequently, as $\R$ \eref{eq:R_factor} is monotonically increasing in $|\rhoE{LL}{RR}|$, the $\R$-factor of any separable two-qubit state obeys
    \begin{equation}
        \R \leq \R_{\mrm{sep}} \coloneqq \frac{X + \frac{1}{2}Y}{X - \frac{1}{2}Y}.
        \label{eq:t2_sep_r}
    \end{equation}
\end{theorem}
\begin{proof}
    For a separable state $\rho = \sum_i p_i \rho_i^{(1)} \otimes \rho_i^{(2)}$, the single-qubit bound $|\bra{R}\rho_i^{(\ell)}\ket{L}|\leq\frac{1}{2}$ gives:
    \begin{equation}
        |\rhoE{LL}{RR}| \leq \sum_i p_i |\bra{R} \rho_i^{(1)} \ket{L}|\;| \bra{R} \rho_i^{(2)} \ket{L}|\leq\frac{1}{4}.
        \label{eq:UB_sep}
    \end{equation}
    Substituting \eqnref{eq:UB_sep} into \eqnref{eq:R_factor} yields \eqnref{eq:t2_sep_r}.
\end{proof}

The separable states of \tabref{tab:states_R_examples}, i.e.~the product state $\ket{HV}$, its $C_4$-symmetric mixture with $\ket{VH}$ and its azimuthally averaged counterpart, attain $\R=\Rsep$, while the product state $\ket{RR}$ yields the minimal value $\R=1$. As the set of separable states is convex (hence connected), every $\R\in[1,\Rsep]$ is attained by some separable state, whereas any $\R\in(\Rsep,\R_{\max}]$ certifies entanglement. Hence, $\R$ constitutes an entanglement witness~\cite{Guehne2009} with a tight threshold $\Rsep$, i.e.~it is optimal given access only to the $\Delta\varphi$-statistics~\eref{eq:av_cross_section}.

\prlparagraph{Optimality of $\R$ as an entanglement witness}%
Beyond the $\Delta\varphi$-statistics $p(\Delta\varphi|\theta_1,\theta_2)$, two distinct sources of suboptimality must be considered. First, the Compton POVM \eref{eq:POVM_element} is not informationally complete~\cite{Prugovecki1977,Busch1991,Renes2004,DAriano2005}, so that some entangled states cannot be certified by \emph{any} analysis of Compton-scattering data; second, even when they can, $\R$ need not be the optimal witness once the full joint distribution $p(\varphi_1,\varphi_2|\theta_1,\theta_2)$ is considered.

As the effective scattering operators \eref{eq:Sigma_l} contain no $\sz$ in the circular basis \eref{eq:RL_basis}, any linear entanglement witness~\cite{Guehne2009} that can be constructed with the Compton POVM \eref{eq:POVM_element} at fixed polar angles is a combination of the nine operators $\s{\mu}\otimes\s{\nu}$ with $\mu,\nu\in\{0,x,y\}$ (see \appref{app:witnesses}). In particular, the Compton-scattering data contain no information about the diagonal entries of $\rho$ in \eqnref{eq:gen_mixed} or, for instance, the difference $\rhoE{LR}{RR}-\rhoE{LL}{RL}$ entering $\langle\sx\otimes\sz\rangle$.

Restricting further to LAS states, whose single-photon expectation values $\langle\s{i}\otimes\1\rangle$ and $\langle\1\otimes\s{j}\rangle$ ($i,j\in\{x,y\}$) vanish, we may also restrict, without loss of generality, to witnesses built solely on the linear-polarisation correlations $\s{i}\otimes\s{j}$ (see \appref{app:witnesses}). The optimal ones then form two one-parameter branches. The first, $W_\chi\coloneqq\frac14\1+\frac12\left(\ee^{-\ii\chi}\ketbra{RR}{LL}+\mathrm{h.c.}\right)$, yields $\langle W_\chi\rangle_\rho=\tfrac14+\Re{\ee^{-\ii\chi}{\rhoE{LL}{RR}}^*}$, whose minimum over $\chi$, $\tfrac14-|\rhoE{LL}{RR}|$, is monotonic in $\R$, so that $\R$ effectively implements $W_\chi$ optimised over $\chi$. The second, obtained by replacing $\ketbra{RR}{LL}$ with $\ketbra{RL}{LR}$, is the analogue of $\R$ for $\rhoE{LR}{RL}$ and is optimal whenever $|\rhoE{LR}{RL}|>|\rhoE{LL}{RR}|$, but requires access to $\varphi_1+\varphi_2$ in \eqnref{eq:gen_cross_section}, i.e.~to the full $p(\varphi_1,\varphi_2|\theta_1,\theta_2)$. We summarise these results, proven in \appref{app:witnesses}, as follows:
\begin{theorem}[Optimality of $\R$]
\label{thm:R_optimal}
Let $\rho$ in \eqnref{eq:gen_mixed} be LAS with $|\rhoE{LL}{RR}|\ge|\rhoE{LR}{RL}|$. Then, the criterion $\R>\Rsep$, i.e.~$|\rhoE{LL}{RR}|>1/4$, is equivalent to the optimal linear entanglement witness that can be constructed with the Compton POVM \eref{eq:POVM_element}. Moreover, if $\R\le\Rsep$, no other analysis of Compton-scattering data can certify that $\rho$ is entangled.
\end{theorem}

For non-LAS states, in which at least one photon is (partially) linearly polarised, the local terms of a witness may improve the detection, so that, for some states, entanglement can be certified by Compton scattering even when $\max\{|\rhoE{LL}{RR}|,|\rhoE{LR}{RL}|\}\le1/4$. The optimal witness can then be determined numerically by semidefinite programming over the nine accessible expectation values~\cite{czapla2026inprep}.

To illustrate the limits of the two branches in certifying entanglement even for LAS states, we compare them with the negativity, $\mathcal{N}(\rho)\coloneqq\frac{\Vert\rho^{\Gamma}\Vert_1-1}{2}$~\cite{Horodecki2009}, where $\rho^\Gamma$ denotes the partial transpose of $\rho$, i.e.~$(\ketbra{ij}{kl})^\Gamma\coloneqq\ketbra{il}{kj}$ for $i,j,k,l\in\{R,L\}$. For two qubits, the negativity constitutes a faithful entanglement measure, i.e.~$\mathcal{N}(\rho)>0$ \emph{iff} $\rho$ is entangled, by virtue of the Peres--Horodecki criterion~\cite{Peres1996,Horodecki1996}. We consider the two-parameter family of states ($0\le p\le1$ and $0\le\alpha\le\pi/2$)
\begin{equation}
    \rho_{p,\alpha} \coloneqq p\,\proj{\Xi_\alpha} + \frac{1-p}{4}\1,
    \label{eq:state_p_alpha}
\end{equation}
where $\ket{\Xi_\alpha}\coloneqq\cos\alpha\ket{\Phi_-}+\ii\sin\alpha\ket{\Psi_-}$ with $\ket{\Psi_-}\coloneqq\tfrac{1}{\sqrt{2}}\left(\ket{RL}-\ket{LR}\right)$. Note that the state $\rho_{p,\alpha}$ is LAS but not $C_4$-symmetric for $p>0$ and $0<\alpha<\pi/2$, with anti-diagonal coherences $|\rhoE{LL}{RR}|=\tfrac{p}{2}\cos^2\!\alpha$ and $|\rhoE{LR}{RL}|=\tfrac{p}{2}\sin^2\!\alpha$. However, $\ket{\Xi_\alpha}$ is maximally entangled for any $\alpha$, so that $\mathcal{N}(\rho_{p,\alpha})=\max\{0,(3p-1)/4\}$.

\begin{figure}[t!]
  \centering
  \includegraphics[width=0.85\linewidth]{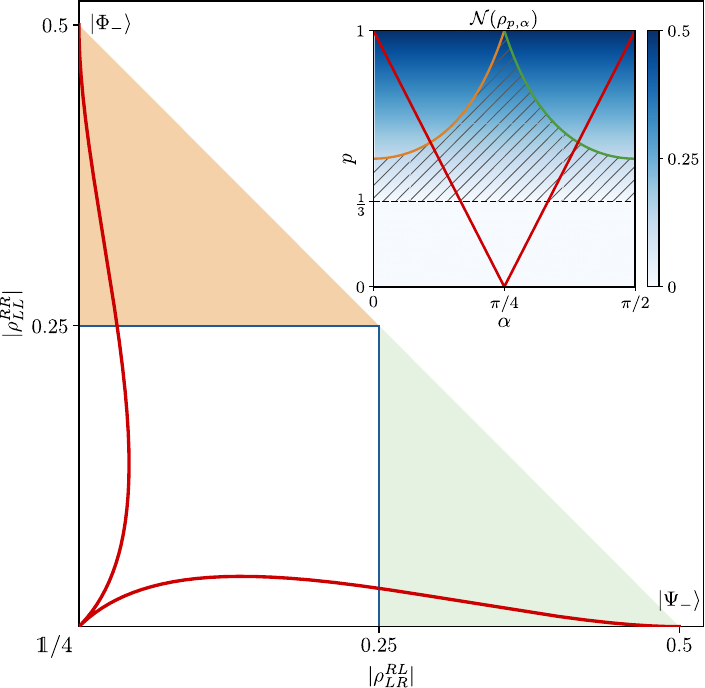}
  \caption{
  \textbf{Moduli of the anti-diagonal coherences} of $\rho_{p,\alpha}$ defined in \eqnref{eq:state_p_alpha} along the curve $p(\alpha)=|1-4\alpha/\pi|$ (\emph{solid red}) within the physical region $|\rhoE{LL}{RR}| + |\rhoE{LR}{RL}| \le \tfrac12$; coloured regions represent regimes in which the first (orange) or second (green) branch of optimal witnesses certifies entanglement, i.e.~$|\rhoE{LL}{RR}|>1/4$ or $|\rhoE{LR}{RL}|>1/4$. \emph{Inset}: negativity $\mathcal{N}(\rho_{p,\alpha})=\max\{0,(3p-1)/4\}$ over the $(p,\alpha)$ plane, with the thresholds $|\rhoE{LL}{RR}|=1/4$ (orange) and $|\rhoE{LR}{RL}|=1/4$ (green) and the curve $p(\alpha)$ marked; in the hatched region the state is entangled (maximally at $p=1$, $\alpha=\tfrac{\pi}{4}$), yet its entanglement cannot be certified from Compton-scattering data.
  }
  \label{fig:state_trajectory}
\end{figure}

In \figref{fig:state_trajectory}, we show both anti-diagonal coherences along the curve $p(\alpha)=|1-4\alpha/\pi|$, which interpolates between $\ket{\Phi_-}$ at $\alpha=0$ and $\ket{\Psi_-}$ at $\alpha=\pi/2$ via the maximally mixed state at $\alpha=\pi/4$. The orange (green) shading marks the regime in which the first (second) branch, optimal for $\alpha\le\pi/4$ ($\alpha\ge\pi/4$), certifies entanglement, i.e.~$|\rhoE{LL}{RR}|>1/4$ ($|\rhoE{LR}{RL}|>1/4$). As shown in the inset, neither branch detects the entanglement of $\rho_{p,\alpha}$ whenever $1/3<p\le1/(2\max\{\cos^2\alpha,\sin^2\alpha\})$. Crucially, as these states are LAS, this entanglement cannot be certified by \emph{any} analysis of Compton-scattering data (see \appref{app:witnesses}), even at $p=1$ and $\alpha=\pi/4$ where the state is maximally entangled.

%%%%%%%%%%%%%%%%%%%%%%%%%%%%%%%%%%%%%%%%%%%%%%%%%%%%%%%%%%%%%%%%%%%%%%%%%%%%%%%%%%%%%%%%%%%
%%%%%%%%%%%%%%%%%%%%%%%%%%%%%%%%%%%%%%%%%%%%%%%%%%%%%%%%%%%%%%%%%%%%%%%%%%%%%%%%%%%%%%%%%%%
\prlparagraph{Quantifying entanglement with the help of $\R$}%
Whenever $\R$ certifies entanglement, whether optimally or not, it also lower-bounds the negativity $\mathcal{N}$. Specifically, for any $\rho$ belonging to a given class $\mathcal{S}$ of states~\eref{eq:gen_mixed},
\begin{equation}
    \mathcal{N}(\rho)\ge\mathcal{N}^{\downarrow}_{\mathcal{S}}(\R)\coloneqq\min_{\substack{\varrho\in\mathcal{S}\\ \R(\varrho)=\R}}\mathcal{N}(\varrho),
    \label{eq:N_lower}
\end{equation}
with $\R\equiv\R(\rho)$. Moreover, if $\mathcal{S}$ is closed under the LOCC map $\mathcal{E}(\rho)\coloneqq\frac12\rho+\frac12(\sz\otimes\sz)\rho(\sz\otimes\sz)$~\cite{Chitambar2014,Horodecki2009}, the minimisation in \eqnref{eq:N_lower} may be restricted to the $C_4$-symmetric states $\mathcal{E}(\mathcal{S})\subseteq\mathcal{S}$, as for any $\varrho\in\mathcal{S}$, $\mathcal{E}(\varrho)$ has the same anti-diagonal coherences, hence the same $\R$, but cannot have larger negativity.

\begin{figure}[t]
    \centering
        \includegraphics[width=\linewidth]{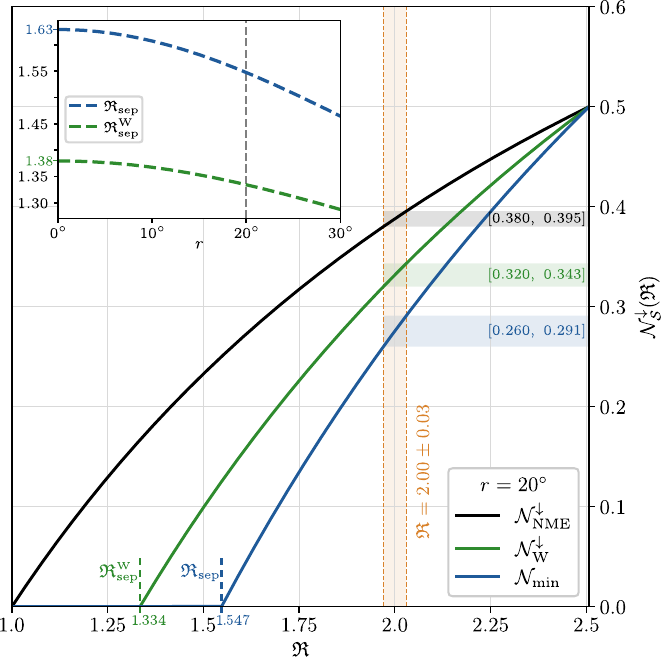}
    \caption{\textbf{Lower bounds on the negativity} implied by the $\R$-factor averaged over a disc of radius $r=20^\circ$ centred at $\theta_1=\theta_2=82^\circ$, assuming NME states $\ket{\Phi_\chi}$ (black), Werner states $\rho_p$ (green) or no assumption on the state, $\mathcal{N}_{\min}$ in \eqnref{eq:nmin} (blue). The vertical dashed blue (green) line marks $\Rsep$ ($\Rsep^W$), above which entanglement is certified for any (Werner) state, and the vertical band the value $\R=2.00\pm0.03$ measured in \citeref{Moskal2025}, which translates into the bounds indicated by the horizontal bands (cf.~\tabref{tab:data}). \emph{Inset}:~$\Rsep$ ($\Rsep^W$) as a decreasing function of the averaging radius $r$, equal to $\approx1.629$ ($\approx1.380$) at $r=0^\circ$ and $\approx1.547$ ($\approx1.334$) at $r=20^\circ$ (vertical grey dashed line; see \appref{app:averaging}).}
    \label{fig:analytical_pred}
\end{figure}

In particular, considering $\mathcal{S}$ to contain all two-qubit states~\eref{eq:gen_mixed}, we obtain the state-independent bound
\begin{equation}
    \mathcal{N}_{\min}(\R)\coloneqq\mathcal{N}^{\downarrow}_{\mathrm{all}}(\R)=\max\left\{0,\kappa-\tfrac12\right\},
    \label{eq:nmin}
\end{equation}
where $\kappa\coloneqq X(\R-1)/[Y(\R+1)]=2|\rhoE{LL}{RR}|\in[0,1]$ by \eqnref{eq:R_factor}. The bound is attained by the GAS states $\kappa\proj{\psi_\xi}+\frac{1-\kappa}{2}\left(\proj{RL}+\proj{LR}\right)$, with $\ket{\psi_\xi}$ defined in \thmref{thm:max_r}. Restricting instead to the \emph{non-maximally entangled} (NME) states $\ket{\Phi_\chi}\coloneqq\sin\frac{\chi}{2}\ket{RR}+\cos\frac{\chi}{2}\ket{LL}$ or the \emph{Werner} states $\rho_p\coloneqq p\proj{\Phi_-}+\frac{1-p}{4}\1$, both $C_4$-symmetric, yields $\mathcal{N}^{\downarrow}_{\mathrm{NME}}=\kappa/2$ and $\mathcal{N}^{\downarrow}_{\mathrm{W}}=\max\{0,(3\kappa-1)/4\}$, respectively. Hence, within these classes, entanglement is inferred for $\R>1$ (NME), $\R> \Rsep^{W}\coloneqq(3X+Y)/(3X-Y)$ (Werner) and $\R>\Rsep = (2X + Y)/(2X - Y)$ (all states). This illustrates how prior knowledge of the class may improve the detection and the inferred amount of entanglement.

These bounds allow us to reinterpret the recent measurement of the J-PET collaboration~\cite{Moskal2025}, in which $\R$ was determined for electron--positron annihilation photons with a plastic-scintillator PET scanner upon averaging over polar angles $\vec{\theta}\coloneqq(\theta_1,\theta_2)\in\Theta$, a disc of radius $r=20^\circ$ centred at $\theta_1=\theta_2=\theta_\mrm{c}$, with either $\theta_\mrm{c}=82^\circ$ or $94^\circ$. As shown in \appref{app:averaging}, such averaging is incorporated by replacing $X$ and $Y$ with their integrals $\bar{X}$ and $\bar{Y}$ over $\Theta$, weighted by the detection efficiency $\eta(\vec{\theta})$. This does not affect the bounds $\mathcal{N}^{\downarrow}_{\mathcal{S}}(\kappa)$, as $\kappa=2|\rhoE{LL}{RR}|$ is a property of the state, whereas the map $\R\mapsto\kappa$ and, hence, the thresholds $\R_{\max}$ and $\Rsep$ are modified (see the inset of \figref{fig:analytical_pred}). For $\theta_\mrm{c}=82^\circ$ ($94^\circ$), averaging lowers $\Rsep$ to $\approx1.55$ ($1.48$) and $\R_{\max}$ to $\approx2.51$ ($2.25$), consistently with the simulations of \citeref{Moskal2025}.

\begin{table}[t]
  \setlength{\tabcolsep}{1.5pt}
  \begin{ruledtabular}
  \begin{tabular}{cccc}
   & & \multicolumn{2}{c}{\textbf{Lower bound} $\mathcal{N}_{\min}$} \\ %[0.4ex]
  \cmidrule(l{0pt}r{1.5pt}){3-4}
  $\theta_\mrm{c}$ & $\R$ & uniform $\eta(\vec{\theta})$ & arbitrary $\eta(\vec{\theta})$ \\
  \hline
  \rule{0pt}{2.8ex}$82^\circ$ & $2.00 \pm 0.03$ & $[0.260, 0.291]$ & $[0.182, 0.210]$ \\
  $94^\circ$ & $1.93 \pm 0.03$ & $[0.306, 0.343]$ & $[0.148, 0.178]$ \\
  \end{tabular}
  \end{ruledtabular}
  \caption{\textbf{Lower bounds on the negativity}, $\mathcal{N}_{\min}$ \eref{eq:nmin}, implied by the $\R$-factors measured (within one standard deviation) in \citeref{Moskal2025} for two subsets of events of the same experiment, selected within discs of radius $r=20^\circ$ centred at $\theta_1=\theta_2=\theta_\mrm{c}$. The detection efficiency within each disc, $\eta(\vec{\theta})$, is assumed independent of the azimuthal angles. The choice $\theta_\mrm{c}=94^\circ$ ($82^\circ$) provides a tighter bound, valid for both event subsets as they refer to the same two-photon state, for a uniform (arbitrary) $\eta(\vec{\theta})$.}
  \label{tab:data}
\end{table}

\figref{fig:analytical_pred} shows the lower bounds on negativity $\mathcal{N}^{\downarrow}_{\mathcal{S}}$ for $\mathcal{S}\in\{\mrm{all},\mrm{W}, \mrm{NME}\}$ as functions of $\R$ for $\theta_\mrm{c}=82^\circ$ assuming uniform $\eta(\vec{\theta})$, while \tabref{tab:data} lists $\mathcal{N}_{\min}\equiv\mathcal{N}^{\downarrow}_{\mrm{all}}$ for $\R$-values measured in \citeref{Moskal2025} for both $\theta_\mrm{c}$, assuming uniform or arbitrary $\eta(\vec{\theta})$. For both $\theta_\mrm{c}$-choices, $\R>\Rsep$, so that entanglement is certified without any assumption on the state or its source, with $\mathcal{N}_{\min}$ quantifying the minimal negativity consistent with the data. Making $\mathcal{N}_{\min}$ independent of $\eta(\vec{\theta})$, by replacing $\bar{Y}/\bar{X}$ with the maximal $Y/X$ within $\Theta$ (see \appref{app:averaging}), lowers its value and favours the $\theta_\mrm{c}=82^\circ$ data, whereas $\theta_\mrm{c}=94^\circ$ is superior for uniform $\eta(\vec{\theta})$. While the results of \citeref{Moskal2025} were presented as evidence of ``non-maximal entanglement'', assuming the photons to be in an NME state $\ket{\Phi_\chi}$ would significantly increase the negativity prediction (cf.~\figref{fig:analytical_pred}).

%%%%%%%%%%%%%%%%%%%%%%%%%%%%%%%%%%%%%%%%%%%%%%%%%%%%%%%%%%%%%%%%%%%%%%%%%%%%%%%%%%%%%%%%%%%
%%%%%%%%%%%%%%%%%%%%%%%%%%%%%%%%%%%%%%%%%%%%%%%%%%%%%%%%%%%%%%%%%%%%%%%%%%%%%%%%%%%%%%%%%%%
\prlparagraph{Conclusions and outlook}
\label{sec:conclusions}
We have shown that the $\R$-factor, the standard benchmark of annihilation-photon experiments since the 1940s, constitutes a rigorous entanglement witness for \emph{any} two-photon polarisation state. Identifying a hierarchy of rotational symmetries, $\mathrm{LAS}\supset C_4\supset\mathrm{GAS}$, we have determined when this certification is optimal. Moreover, we have used $\R$ to lower-bound the negativity, which allowed us to quantify, in a state-independent manner, the entanglement present in the recent J-PET data~\cite{Moskal2025}. Our results can be naturally extended to the full Compton-scattering data, also beyond LAS states, by means of semidefinite programming~\cite{czapla2026inprep}. However, as they rely on the QED description of Compton scattering, entanglement certification that does not depend on the measurement model~\cite{Branciard2013} is desirable, e.g.~by resorting to Fisher-information-based witnesses employed in quantum metrology~\cite{Pezze2009,Strobel2014}. Extending the present framework to the three-photon decays of ortho-positronium~\cite{Acin2001,Hiesmayr2017,Dulski2021,Moskal2021} also remains an open question.

\begin{acknowledgments}
We thank Pawe\l{} Moskal, Sushil Sharma, Sebastiano Mariazzi, Peter Caradonna, Karol \.{Z}yczkowski and Ryszard Horodecki for helpful discussions. This work was supported by the National Science Centre, Poland under grant no. 2023/50/E/ST2/00457 [PC, JK] and by the Polish Ministry of Education and Science under agreement no. 2022/WK/06 [TS]. Claude (Anthropic; Fable~5.2 and Opus~5.5) was used to edit the text and to check derivations, proofs and numerical values; all output was verified by the authors.
\end{acknowledgments}
%%%%%%%%%%%%%%%%%%%%%%%%%%%%%%%%%%%%%%%%%%%%%%%%%%%%%%%%%%%%%%%%%%%%%%%%%%%%%%%%%%%%%%%%%%%

%%%%%%%%%%%%%%%%%%%%%%%%%%%%%%%%%%%%%%%%%%%%%%%%%%%%%%%%%%%%%%%%%%%%%%%%%%%%%%%%%%%%%%%%
\bibliographystyle{myapsrev4-2}
\bibliography{JPET_ent}

%%%%%%%%%%%%%%%%%%%%%%%%%%%%%%%%%%%%%%%%%%%%%%%%%%%%%%%%%%%%%%%%%%%%%%%%%%%%%%%%%%%%%%%%%%%
%APPENDIX
\onecolumngrid
\newpage
\appendix
\part*{Appendices}
%%%%%%%%%%%%%%%%%%%%%%%%%%%%%%%%%%%%%%%%%%%%%%%%%%%%%%%%%%%%%%%%%%%%%%%%%%%%%%%%%%%%%%%%%%%

\section{Derivation of the scattering operator for two Compton-scattered photons}
\label{app:scat_deriv}
\begin{figure}[ht]
    \centering
    \begin{minipage}{0.45\textwidth}
        \centering
        \begin{tikzpicture}
            \begin{feynman}
                % Definicja centralnych wierzchołków dla kanału s
                \vertex (w1);
                \vertex [right=1.75cm of w1] (w2);

                % Wierzchołki wejściowe i wyjściowe
                \vertex [above left=1.2cm of w1] (k) {\(k\)};
                \vertex [below left=1.2cm of w1] (p) {\(p\)};
                \vertex [above right=1.2cm of w2] (kp) {\(k'\)};
                \vertex [below right=1.2cm of w2] (pp) {\(p'\)};

                \diagram* {
                    (p) -- [fermion] (w1),
                    (k) -- [photon] (w1),
                    (w1) -- [fermion, edge label=\(p+k\)] (w2),
                    (w2) -- [photon] (kp),
                    (w2) -- [fermion] (pp),
                };
            \end{feynman}
        \end{tikzpicture}
    \end{minipage}
    \hfill
    \begin{minipage}{0.45\textwidth}
        \centering
        \begin{tikzpicture}
            \begin{feynman}
                % Definicja centralnych wierzchołków dla kanału u
                \vertex (w1);
                \vertex [right=1.75cm of w1] (w2);

                % Wierzchołki wejściowe i wyjściowe
                \vertex [above left=1.2cm of w1] (k) {\(k\)};
                \vertex [below left=1.2cm of w1] (p) {\(p\)};
                \vertex [above right=1.2cm of w2] (kp) {\(k'\)};
                \vertex [below right=1.2cm of w2] (pp) {\(p'\)};

                \diagram* {
                    (p) -- [fermion] (w1),
                    (k) -- [photon] (w2),
                    (w1) -- [fermion, edge label'=\(p-k'\)] (w2),
                    (w1) -- [photon] (kp),
                    (w2) -- [fermion] (pp),
                };
            \end{feynman}
        \end{tikzpicture}
    \end{minipage}
    \caption{$\mathcal{S}$-channel (left) and $\mathcal{U}$-channel (right) Feynman diagrams for Compton scattering.}
    \label{fig:compton}
\end{figure}
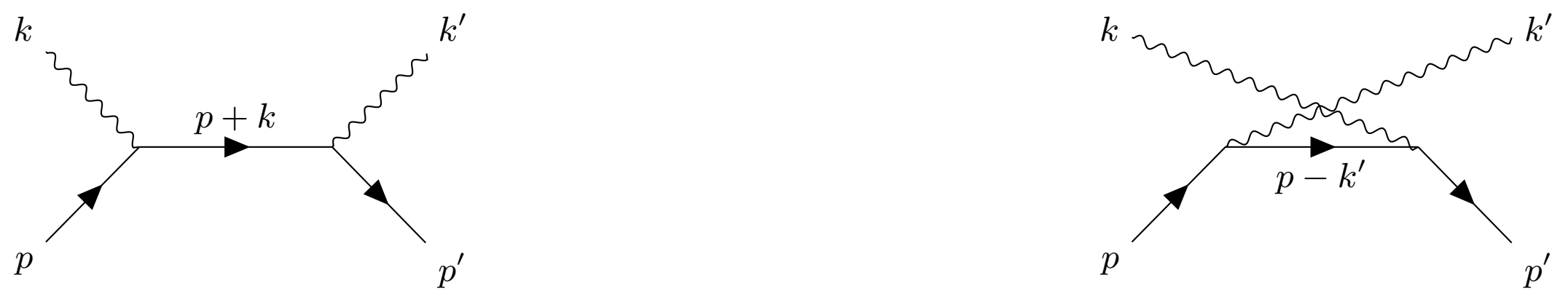

Before detailing the explicit evaluation of the matrix elements, it is crucial to establish the perturbative order of the calculation and the physical justification for computing the off-diagonal interference terms. The scattering process is evaluated strictly at the tree level, which represents the leading-order contribution of $\mathcal{O}(\alpha^2)$ (equivalently $\mathcal{O}(e^4)$) to the bilinear scattering operator. At this order, the dynamics are governed exclusively by the $\mathcal{S}$- and $\mathcal{U}$-channel Compton diagrams~\cite{Peskin:1995}, see \fref{fig:compton}, while higher-order radiative corrections, suppressed by a further factor of $\alpha$, are neglected.

The following notation for each photon labelled by $\ell \in \{1, 2\}$ will be used throughout the calculations:
\begin{itemize}
    \item $a$ will denote a contravariant four-vector.
    \item $a \cdot b = g_{\mu\nu}a^\mu b^\nu$ will explicitly denote the dot product (contraction) of two four-vectors with respect to the spacetime metric $g_{\mu\nu}$ ($+,-,-,-$).
    \item $\slashed a$ is a vector denoted with the Feynman slash notation, defined as $\slashed a = \gamma^\mu a_\mu$, where $\gamma^\mu$ are the Dirac matrices.
    \item $i, j \in \{R, L\}$ denote the incoming circular-polarisation states for the density matrix components.
    \item $\epsilon^\ell_i$ will be the incoming polarisation four-vector for state $i$.
    \item $\epsilon'_\ell$ will be the outgoing polarisation four-vector.
    \item $u_r(p)$ will be the spinor of a fermion with spin $r$ and momentum $p$.
    \item $\bar{\psi}$ will be the Dirac adjoint of a spinor defined as $\bar{\psi}\coloneqq \psi^{\dagger}\gamma^0$.
    \item $m$ is the mass of an electron, $e$ is the elementary charge and $r_e$ is the classical radius of the electron.
\end{itemize}

From the Feynman diagrams, one can easily construct both matrix elements:
\begin{equation}
    \mathcal{M}_{\mathcal{S}}(p',s';k',\epsilon'| p, s; k, \epsilon) = \bar{u}_{s'}(p') \left(\ii e(\slashed{\epsilon}')^*\right) \frac{\ii (\slashed{p}+\slashed{k}+m)}{(p+k)^2-m^2} \left(\ii e\slashed{\epsilon}\right) u_s(p)
\end{equation}
and

\begin{equation}
    \mathcal{M}_{\mathcal{U}}(p',s';k',\epsilon'| p, s; k, \epsilon) = \bar{u}_{s'}(p') \left(\ii e\slashed{\epsilon}\right) \frac{\ii (\slashed{p}-\slashed{k}'+m)}{(p-k')^2-m^2} \left(\ii e(\slashed{\epsilon}')^*\right) u_{s}(p).
\end{equation}

The total scattering amplitude is given by the sum of the amplitudes corresponding to the two channels, $\mathcal{M}=\mathcal{M}_{\mathcal{S}}+\mathcal{M}_{\mathcal{U}}$, which after further simplifications (using the on-shell conditions and transversality, $p\cdot\epsilon=k\cdot\epsilon=p\cdot\epsilon'=k'\cdot\epsilon'=0$) reads:
\begin{equation}
    \mathcal{M} (p',s';k',\epsilon'| p, s; k, \epsilon)
    = \ii e^2\bar{u}_{s'}(p') \left(\frac{(\slashed\epsilon')^* \slashed \epsilon \slashed k}{2p_\alpha k^\alpha} + \frac{\slashed \epsilon (\slashed \epsilon')^* \slashed k'}{2p_\alpha k'^\alpha}\right) u_s(p).
    \label{eq:Mdupa}
\end{equation}

Now, we are interested in finding the differential scattering cross-section of a photon whose state is expressed explicitly in the polarisation (here, $\{R,L\}$) basis. Hence, we express it as follows:
\begin{equation}
    \frac{\dd \sigma}{\dd \Omega} = \tr{\varrho\, \hat{\Sigma}(\Omega)}= \sum_{i,j\in\{R,L\}} \!\!\! \varrho^i_j \; [\hat{\Sigma}(\Omega)]_{i}^j,
\end{equation}
where $\varrho^{i}_j=\bra{i}\varrho\ket{j}$ is the matrix element of the incoming (single) photon density matrix, whereas $[\hat{\Sigma}(\Omega)]^j_i=\bra{j}\hat{\Sigma}\ket{i}$ is the 
element of the effective scattering operator. The latter can be expressed with help of \eqnref{eq:Mdupa} by fixing the input polarisation to the circular basis, $\epsilon_i$ with $i\in \{R,L\}$, as follows
\begin{equation}
    [\hat{\Sigma}(\Omega)]_{i}^j =  \frac{1}{64\pi^2m^2}
    \left(\frac{k'^0}{k^0}\right)^2
    \sum_{s'} \sum_{s,\tilde{s}} p(s,\tilde{s})
    \mathcal{M} (p',s';k',\epsilon'| p, s; k, \epsilon_i)
    \mathcal{M} (p',s';k',\epsilon'| p, \tilde{s}; k, \epsilon_j)^*,
\end{equation}
where we need to sum over the unobserved final electron spin states ($s'$) and average over their initial distribution, assumed unpolarised, $p(s,\tilde{s})=\tfrac{1}{2}\delta_{s\tilde{s}}$.

We derive the explicit form of $\hat{\Sigma}$ following standard rules of Feynman algebra and arrive at: 
\begin{equation}
    [\hat{\Sigma}(\Omega)]_{i}^j 
    = \frac{r_e^2}{32}
    \left(\frac{k'^0}{k^0}\right)^2 \left[ \frac{\mathrm{{\bf I}}}{(p\cdot k)^2} + \frac{\mathrm{{\bf II}}}{(p\cdot k)(p\cdot k')} + \frac{\mathrm{{\bf III}}}{(p\cdot k')(p\cdot k)} + \frac{\mathrm{{\bf IV}}}{(p\cdot k')^2} \right],
\end{equation}
with:
\begin{align}
    \mathrm{{\bf I}}   & = \mathrm{Tr}[(\slashed \epsilon')^* \slashed \epsilon_i \slashed k (\slashed p + m) \slashed k (\slashed \epsilon_j)^*\slashed \epsilon' (\slashed p' + m)], \\
    \mathrm{{\bf II}}  & = \mathrm{Tr}[(\slashed \epsilon')^* \slashed \epsilon_i \slashed k(\slashed p+m)\slashed k' \slashed \epsilon' (\slashed \epsilon_j)^* (\slashed p'+m)],      \\
    \mathrm{{\bf III}} & = \mathrm{Tr}[\slashed \epsilon_i (\slashed \epsilon')^* \slashed k'(\slashed p+m)\slashed k (\slashed \epsilon_j)^*\slashed \epsilon' (\slashed p'+m)],      \\
    \mathrm{{\bf IV}}  & = \mathrm{Tr}[\slashed \epsilon_i (\slashed \epsilon')^* \slashed k'(\slashed p+m)\slashed k' \slashed \epsilon' (\slashed \epsilon_j)^* (\slashed p'+m)],
\end{align}
which finally simplifies to:
\begin{equation}
        [\hat{\Sigma}(\Omega)]_{i}^j =    \frac{r_e^2}{4}
    \left(\frac{k'^0}{k^0}\right)^2\left[ \delta_{ij}\left(\frac{k'^0}{k^0} + \frac{k^0}{k'^0} - 2 \right) + 4\,(\epsilon_i \cdot \epsilon')((\epsilon_j)^{*} \cdot \epsilon') \right].
    \label{eq:Mij-real}
\end{equation}

Summing over the two unobserved final polarisations $\epsilon'$ and dropping an overall factor of $2$, which cancels upon normalisation [cf.~$\mathcal{A}=\sigma_\mathrm{KN}/(8\pi)$ in \appref{app:witnesses}], yields \eqnref{eq:Sigma_l} with $A(\theta)$ and $B(\theta)$ given in \eqnref{eq:AB}. Because the ratio $\frac{k'^0}{k^0}$ depends solely on the photon's scattering solid angle $\Omega$, the scattering operator $\hat{\Sigma}$ likewise depends only on $\Omega$. Denoting the input product state of the photon in polarisation $i$ and the electron with spin $s$ as $\ket{i;s}\equiv\ket{i}_{\gamma}\ket{s}_{e}$, and the general output state of the photon scattered into the solid angle $\Omega$ and the electron possessing spin $s'$ by $\ket{\Omega;s'}\equiv\ket{\Omega;s'}_{\gamma,e}$, we may rewrite \eqnref{eq:Mij-real} as:
\begin{equation}
    [\hat{\Sigma} (\Omega)]_{i}^j=\frac{1}{128\pi^2m^2}\left(\frac{k'^0}{k^0}\right)^2\sum_{ss'} \bra{\Omega;s'}\hat{T}\ket{i;s} \bra{j;s} \hat{T}^\dagger\ket{\Omega;s'}.
\end{equation}
Moreover, such notation allows us to identify the effective scattering operator as
\begin{equation}
    \hat{\Sigma} (\Omega) = \frac{1}{128\pi^2m^2}\left(\frac{k'^0}{k^0}\right)^2 \sum_{ss'} \bra{s} \hat{T}^\dagger\ket{\Omega;s'} \bra{\Omega;s'}\hat{T}\ket{s},
    \label{eq:Sigma}
\end{equation}
which acts on the Hilbert space defined by the polarisation degrees of freedom for the incoming photon. The explicit form of $\hat{\Sigma}(\Omega)$ written in the $i\in\{R,L\}$ circular basis is stated in \eqnref{eq:Sigma_l} of the main text. 

Considering now the joint Compton scattering of two photons off distinct electrons, the resulting $\hat{T}$-matrix in \eqnref{eq:Sigma} exhibits a tensor-product structure, $\hat{T}_{12}=\hat{T}_1\otimes\hat{T}_2$, which is then naturally inherited by the effective two-photon scattering operator, i.e.:
\begin{equation}
    \hat{\Sigma}_{12}(\vec{\Omega}) = \hat{\Sigma}_1(\Omega_1) \otimes \hat{\Sigma}_2(\Omega_2)^*,
    \label{eq:Sigma_12_tensor}
\end{equation}
where the complex conjugation reflects that photon 2 propagates along $-z$, so that its azimuthal angle enters with the opposite sign. Writing out \eqnref{eq:Sigma_12_tensor} explicitly in the polarisation basis, we obtain
\begin{equation}
\hat{\Sigma}_{12}(\vec{\Omega}) =  
    \begin{pmatrix}
        X & A(\theta_1) B(\theta_2) \ee ^{-2\mathrm{i}\varphi_2} & A(\theta_2) B(\theta_1) \ee^{2\mathrm{i}\varphi_1} & Y \ee^{2\mathrm{i}(\varphi_1 - \varphi_2)} \\
        A(\theta_1) B(\theta_2) \ee^{2\mathrm{i}\varphi_2} & X & Y \ee^{2\mathrm{i}(\varphi_1 + \varphi_2)} & A(\theta_2) B(\theta_1) \ee^{2\mathrm{i}\varphi_1} \\
        A(\theta_2) B(\theta_1) \ee^{-2\mathrm{i}\varphi_1} & Y \ee^{-2\mathrm{i}(\varphi_1 + \varphi_2)} & X & A(\theta_1) B(\theta_2) \ee^{-2\mathrm{i}\varphi_2} \\
         Y \ee^{-2\mathrm{i}(\varphi_1 - \varphi_2)} & A(\theta_2) B(\theta_1) \ee^{-2\mathrm{i}\varphi_1} & A(\theta_1) B(\theta_2) \ee^{2\mathrm{i}\varphi_2} & X
    \end{pmatrix},
    \label{eq:Sigma12_matrix}
\end{equation}
where $X \coloneqq A(\theta_1) A(\theta_2)$ and $Y \coloneqq B(\theta_1) B(\theta_2)$, with $A(\theta_\ell)$ and $B(\theta_\ell)$ given in \eqnref{eq:AB}.

%%%%%%%%%%%%%%%%%%%%%%%%%%%%%%%%%%%%%%%%%%%%%%%%%%%%%%%%%%%%%%%%%%%%%%%%%%%%%%%%%%%%%%%%%%%%%%%%%%%%%%%%%%%%%%%%%%%%%%%%%%%%%%%%%
\section{Entanglement witnesses accessible with Compton scattering}
\label{app:witnesses}

\subsection*{Bloch form of the conditional single-photon POVM}
Defining the solid-angle measure as $\frac{\dd\Omega_\ell}{4\pi}=\dd\mu_{\theta_\ell}\frac{\dd\varphi_\ell}{2\pi}$ with $\dd\mu_\theta\coloneqq\frac12\sin\theta\,\dd\theta$, the probability density of the scattering directions $\vec{\Omega}=(\Omega_1,\Omega_2)$ with respect to it follows from \eqnref{eq:psi_cross} upon normalisation,
\begin{equation}
    p(\vec{\Omega})=\frac{1}{\mathcal{A}^2}\,\tr{\rho\,\hat{\Sigma}_1(\Omega_1)\otimes\hat{\Sigma}_2(\Omega_2)^*},
    \label{eq:p_Omega}
\end{equation}
where $\mathcal{A}\coloneqq\int_0^\pi\!\dd\mu_\theta\,A(\theta)=\frac{r_e^2}{72}\left(40-27\ln3\right)=\sigma_\mathrm{KN}/(8\pi)$ follows from integrating \eqnref{eq:AB}, with $\sigma_\mathrm{KN}$ being the total Klein--Nishina cross-section at $511\,$keV~\cite{Peskin:1995}. As $\sigma_\mathrm{KN}$ is polarisation-independent, the probability that a photon is Compton-scattered does not depend on its state. Hence, the normalisation in \eqnref{eq:p_Omega}, i.e.~conditioning on both photons being scattered, does \emph{not} constitute post-selection.

As $U_z(\varphi_\ell)$ in \eqnref{eq:Sigma_l} does not depend on $\theta_\ell$, we have $\int_0^{2\pi}\!\frac{\dd\varphi_\ell}{2\pi}\hat{\Sigma}_\ell(\Omega_\ell)=A(\theta_\ell)\1$, so that the polar angles $\vec{\theta}=(\theta_1,\theta_2)$ are distributed as
\begin{equation}
    p(\vec{\theta})=\iint_0^{2\pi}\!\frac{\dd\varphi_1\dd\varphi_2}{(2\pi)^2}\,p(\vec{\Omega})=\frac{A(\theta_1)A(\theta_2)}{\mathcal{A}^2},
    \label{eq:p_theta}
\end{equation}
independently of $\rho$. The polar angles thus act as a measurement setting drawn at random independently of the state, so that conditioning on them does \emph{not} constitute post-selection either. The conditional distribution of the azimuthal angles $\vec{\varphi}=(\varphi_1,\varphi_2)$ then reads
\begin{equation}
    p(\vec{\varphi}|\vec{\theta})=\frac{p(\vec{\Omega})}{p(\vec{\theta})}=\tr{\rho\,\hat{\Sigma}_{\varphi_1|\theta_1}\otimes\hat{\Sigma}^*_{\varphi_2|\theta_2}},
    \label{eq:p_cond}
\end{equation}
which defines the \emph{conditional single-photon POVM} elements $\hat{\Sigma}_{\varphi_\ell|\theta_\ell}\coloneqq\hat{\Sigma}_\ell(\Omega_\ell)/A(\theta_\ell)$, normalised as $\int_0^{2\pi}\!\frac{\dd\varphi_\ell}{2\pi}\hat{\Sigma}_{\varphi_\ell|\theta_\ell}=\1$. Using \eqnref{eq:Sigma_l} and $U_z(\phi)\,\sx\,U_z^\dagger(\phi)=\cos2\phi\,\sx-\sin2\phi\,\sy$, they take the Bloch form
\begin{equation}
    \hat{\Sigma}_{\varphi_\ell|\theta_\ell} = \1 + \frac{B(\theta_\ell)}{A(\theta_\ell)}\left(\cos2\varphi_\ell\,\sx-\sin2\varphi_\ell\,\sy\right),
    \label{eq:Bloch_POVM}
\end{equation}
with the sign of the $\sy$-term reversed for the second photon, which enters via $\hat{\Sigma}^*_{\varphi_2|\theta_2}$. Hence, the Bloch vectors of all the POVM elements lie in the $xy$-plane.

\subsection*{Most general linear witness}
Any operator whose expectation value is a linear functional of the measured distribution \eref{eq:p_cond} is of the form $W[K]=\iint_0^{2\pi}\frac{\dd\varphi_1\dd\varphi_2}{(2\pi)^2}K(\varphi_1,\varphi_2)\,\hat{\Sigma}_{\varphi_1|\theta_1}\otimes\hat{\Sigma}^*_{\varphi_2|\theta_2}$ for some real kernel $K$. By \eqnref{eq:Bloch_POVM}, it lies in the span of $\s{\mu}\otimes\s{\nu}$ with $\mu,\nu\in\{0,x,y\}$, and conversely each such operator is obtained with a product kernel $k^{(1)}_\mu(\varphi_1)k^{(2)}_\nu(\varphi_2)$ built from $k^{(\ell)}_0=1$, $k^{(\ell)}_x=2\tfrac{A(\theta_\ell)}{B(\theta_\ell)}\cos2\varphi_\ell$ and $k^{(1)}_y=-2\tfrac{A(\theta_1)}{B(\theta_1)}\sin2\varphi_1$, $k^{(2)}_y=2\tfrac{A(\theta_2)}{B(\theta_2)}\sin2\varphi_2$. Hence, the most general accessible operator reads
\begin{equation}
    W=\sum_{\mu,\nu\in\{0,x,y\}} w_{\mu\nu}\,\s{\mu}\otimes\s{\nu},
    \label{eq:general_W}
\end{equation}
whereas the seven expectation values involving $\sz$, which determine the diagonal entries of $\rho$ in \eqnref{eq:gen_mixed} as well as differences of its off-diagonal entries such as $\rhoE{LR}{RR}-\rhoE{LL}{RL}$, remain inaccessible. For any product state with Bloch vectors $\vec{a}$, $\vec{b}$, $\langle W\rangle=\tilde{\vec{a}}^\TT w\,\tilde{\vec{b}}$ with $\tilde{\vec{a}}\coloneqq(1,\vec{a}_\perp)^\TT$, $\vec{a}_\perp\coloneqq(a_x,a_y)^\TT$, and similarly $\tilde{\vec{b}}$, so that, by convexity, $W$ is an entanglement witness \emph{iff} $\tilde{\vec{a}}^\TT w\,\tilde{\vec{b}}\ge0$ for all $\vec{a}_\perp$, $\vec{b}_\perp$ in the unit disc.

\begin{figure}[ht]
  \centering
  \includegraphics[width=0.85\textwidth]{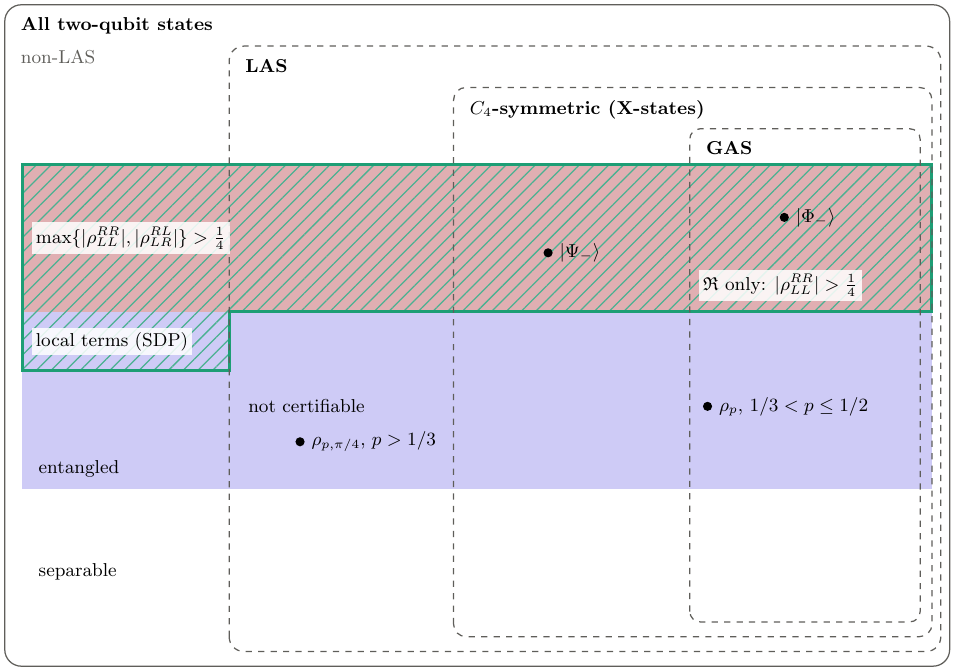}
  \caption{\textbf{Schematic hierarchy of two-photon states and of the entanglement certifiable with Compton scattering}. Dashed boxes mark the nested classes of LAS, $C_4$-symmetric (X-) and GAS states. Purple: entangled states. Hatched (teal outline): states whose entanglement can be certified by some---linear or non-linear---analysis of Compton-scattering data. Coral: states certified by the moduli of the anti-diagonal coherences, $\max\{|\rhoE{LL}{RR}|,|\rhoE{LR}{RL}|\}>1/4$, which reduces to $\R>\Rsep$ whenever $|\rhoE{LL}{RR}|\ge|\rhoE{LR}{RL}|$, in particular for GAS states. Within the LAS class, the two regions coincide (\thmref{thm:R_optimal}), whereas for non-LAS states local terms may certify further entanglement via semidefinite programming. Dots mark the singlet \eref{eq:Phi_-} and $\ket{\Psi_-}$, as well as the entangled states whose entanglement cannot be certified:~the Werner states $\rho_p$ with $1/3<p\le1/2$ and the states $\rho_{p,\pi/4}$ of \eqnref{eq:state_p_alpha} with $p>1/3$. Region sizes are not to scale (TikZ code prepared with the assistance of Claude).}
  \label{fig:compton_venn}
\end{figure}

\subsection*{LAS states}
For LAS states, $\rho=\rho_\trm{X}+(a\,\sz\otimes\sigma_++b\,\sigma_+\otimes\sz+\mathrm{h.c.})$, where $\rho_\trm{X}$ is the X-state obtained from $\rho$ by setting $\rhoE{LR}{RR}$, $\rhoE{LL}{RL}$, $\rhoE{RL}{RR}$ and $\rhoE{LL}{LR}$ to zero, $\sigma_+\coloneqq\ketbra{R}{L}$, $a\coloneqq\rhoE{RL}{RR}$ and $b\coloneqq\rhoE{LR}{RR}$. The second term involves $\sz$ and is thus inaccessible.
Conjugation with $\sz\otimes\sz$ flips the signs of the local terms $\s{i}\otimes\1$, $\1\otimes\s{j}$ ($i,j\in\{x,y\}$), while leaving the correlations $\s{i}\otimes\s{j}$ intact. Hence, as $\langle\s{i}\otimes\1\rangle_\rho=\langle\1\otimes\s{j}\rangle_\rho=0$ for LAS states, averaging any witness \eref{eq:general_W} with its conjugate yields a witness without local terms and with the same expectation value, i.e.~without loss of generality
\begin{equation}
    W_R=\frac14\Big(\1+\sum_{i,j\in\{x,y\}}R_{ij}\,\s{i}\otimes\s{j}\Big),
    \qquad
    \langle W_R\rangle_\rho=\frac14\left(1+\Tr\{R^\TT T\}\right),
    \label{eq:W_R}
\end{equation}
where $T_{ij}\coloneqq\langle\s{i}\otimes\s{j}\rangle_\rho$ is the linear-polarisation correlation matrix. As $\min_{\vec{a}_\perp,\vec{b}_\perp}(1+\vec{a}_\perp^\TT R\,\vec{b}_\perp)=1-\Vert R\Vert_\infty$ over the unit disc, $W_R$ is a witness \emph{iff} $R$ is a contraction. Since $\langle W_R\rangle$ is linear in $R$, its minimum is attained at the extreme points of the set of contractions, i.e.~at orthogonal matrices $R\in\mathrm{O}(2)$, which are either reflections or rotations,
\begin{equation}
    R^{(-)}_\chi=\begin{pmatrix}\cos\chi&\sin\chi\\ \sin\chi&-\cos\chi\end{pmatrix},
    \qquad
    R^{(+)}_\chi=\begin{pmatrix}\cos\chi&-\sin\chi\\ \sin\chi&\cos\chi\end{pmatrix}.
\end{equation}
Substituting these into \eqnref{eq:W_R} yields the two branches
\begin{subequations}
\label{eq:two_branches}
\begin{align}
    W^{(-)}_\chi &= \frac14\1 + \frac12\left(\ee^{-\ii\chi}\ketbra{RR}{LL}+\mathrm{h.c.}\right),\\
    W^{(+)}_\chi &= \frac14\1 + \frac12\left(\ee^{-\ii\chi}\ketbra{RL}{LR}+\mathrm{h.c.}\right),
\end{align}
\end{subequations}
with $\langle W^{(-)}_\chi\rangle=\tfrac14+\Re{\ee^{-\ii\chi}{\rhoE{LL}{RR}}^*}$ and $\langle W^{(+)}_\chi\rangle=\tfrac14+\Re{\ee^{-\ii\chi}{\rhoE{LR}{RL}}^*}$, where the reflection branch $W^{(-)}_\chi$ coincides with $W_\chi$ of the main text, i.e.~the first branch underlying $\R$, and the rotation branch $W^{(+)}_\chi$ with the second one.

Optimising over all contractions, von Neumann's trace inequality gives $\min_R\langle W_R\rangle_\rho=\tfrac14(1-\Vert T\Vert_1)$ with $\Vert T\Vert_1=4\max\{|\rhoE{LL}{RR}|,|\rhoE{LR}{RL}|\}$, so that the optimum is always attained within one of the two branches, namely by the reflection (rotation) branch whenever $\det T<0$ ($\det T>0$), as $\det T=4(|\rhoE{LR}{RL}|^2-|\rhoE{LL}{RR}|^2)$, with both being equally good for $\det T=0$, and entanglement is certified if $\max\{|\rhoE{LL}{RR}|,|\rhoE{LR}{RL}|\}>1/4$.

\subsection*{Optimality}
This criterion is necessary and sufficient, not only among linear witnesses but for any processing of the Compton data. For any LAS state, the Compton data are fully determined by $\rhoE{LL}{RR}$ and $\rhoE{LR}{RL}$, the diagonal entries of $\rho$ being inaccessible. If both satisfy $|\cdot|\le1/4$, the data are reproduced by the X-state
\begin{equation}
    \sigma=\frac14\1+\left(\rhoE{LL}{RR}\ketbra{RR}{LL}+\rhoE{LR}{RL}\ketbra{RL}{LR}+\mathrm{h.c.}\right),
\end{equation}
whose eigenvalues, $\tfrac14\pm|\rhoE{LL}{RR}|$ and $\tfrac14\pm|\rhoE{LR}{RL}|$, are merely interchanged by the partial transposition, so that $\sigma$ is a valid PPT, hence separable, state~\cite{Peres1996,Horodecki1996}. If only $\Delta\varphi$ is recorded, $\rhoE{LR}{RL}$ is lost and the criterion reduces to $|\rhoE{LL}{RR}|>1/4$, i.e.~$\R>\Rsep$. Finally, for non-LAS states, the local terms in \eqnref{eq:general_W} may improve the detection, in which case the optimal witness follows from a semidefinite program over the accessible expectation values~\cite{czapla2026inprep}. The resulting hierarchy is summarised in \figref{fig:compton_venn}.

%%%%%%%%%%%%%%%%%%%%%%%%%%%%%%%%%%%%%%%%%%%%%%%%%%%%%%%%%%%%%%%%%%%%%%%%%%%%%%%%%%%%%%%%%%%%%%%%%%%%%%%%%%%%%%%%%%%%%%%%%%%%%%%%%
\section{Compton scattering POVM averaged over polar angles}
\label{app:averaging}
In experiments, the data are integrated over a range of polar angles, $\vec{\theta}\in\Theta$, e.g.~the disc considered in the main text. By \eqnref{eq:p_theta}, the probability of an event to fall into $\Theta$,
\begin{equation}
    p(\vec{\theta}\in\Theta)=\iint_\Theta\!\dd\mu_{\vec{\theta}}\,p(\vec{\theta})=\frac{\bar{X}}{\mathcal{A}^2},
    \qquad
    \bar{X}\coloneqq\iint_\Theta\!\dd\mu_{\vec{\theta}}\,A(\theta_1)A(\theta_2),
    \qquad
    \bar{Y}\coloneqq\iint_\Theta\!\dd\mu_{\vec{\theta}}\,B(\theta_1)B(\theta_2),
    \label{eq:p_Theta}
\end{equation}
with $\dd\mu_{\vec{\theta}}\coloneqq\dd\mu_{\theta_1}\dd\mu_{\theta_2}$, again does not depend on $\rho$, so that selecting events within $\Theta$ does \emph{not} constitute post-selection either. The azimuthal statistics of the selected events, $p(\vec{\varphi}|\Theta)=\iint_\Theta\dd\mu_{\vec{\theta}}\,p(\vec{\Omega})/p(\vec{\theta}\in\Theta)$, are then described by the \emph{conditional two-photon POVM}
\begin{equation}
    \hat{\Sigma}_{\varphi_1,\varphi_2|\Theta}
    \coloneqq\frac{1}{\bar{X}}\iint_\Theta\!\dd\mu_{\vec{\theta}}\;\hat{\Sigma}_1(\Omega_1)\otimes\hat{\Sigma}_2(\Omega_2)^*
    =\iint_\Theta\!\dd\mu_{\vec{\theta}}\;p(\vec{\theta}|\Theta)\,\hat{\Sigma}_{\varphi_1|\theta_1}\otimes\hat{\Sigma}^*_{\varphi_2|\theta_2},
    \label{eq:POVM_Theta}
\end{equation}
i.e.~a convex mixture of the fixed-angle POVMs with the state-independent weights 
\begin{equation}
  p(\vec{\theta}|\Theta)\coloneqq \frac{p(\vec{\theta})}{p(\vec{\theta}\in\Theta)}
  =\frac{A(\theta_1)A(\theta_2)/\mathcal{A}^2}{\bar{X}/\mathcal{A}^2}
  =\frac{A(\theta_1)A(\theta_2)}{\bar{X}},
\end{equation}
recalling \eqnref{eq:p_theta}. In practice, the selected events are further weighted by the detection efficiency $\eta(\vec{\theta})$. As long as $\eta$ does not depend on the azimuthal angles, it is state-independent, so that the above holds with $\dd\mu_{\vec{\theta}}\to\eta(\vec{\theta})\,\dd\mu_{\vec{\theta}}$. In the main text we assume $\eta$ to be uniform within $\Theta$, while bounds valid for any $\eta$ are discussed below.

Substituting the Bloch form \eref{eq:Bloch_POVM}, \eqnref{eq:POVM_Theta} reads
\begin{equation}
    \hat{\Sigma}_{\varphi_1,\varphi_2|\Theta}
    =\1\otimes\1+\frac{\beta_1}{\bar{X}}\,\hat{n}_1\otimes\1+\frac{\beta_2}{\bar{X}}\,\1\otimes\hat{n}^*_2+\frac{\bar{Y}}{\bar{X}}\,\hat{n}_1\otimes\hat{n}^*_2,
    \label{eq:POVM_Theta_Bloch}
\end{equation}
with $\hat{n}_\ell\coloneqq\cos2\varphi_\ell\,\sx-\sin2\varphi_\ell\,\sy$ and
\begin{equation}
    \beta_1\coloneqq\iint_\Theta\!\dd\mu_{\vec{\theta}}\,B(\theta_1)A(\theta_2),
    \qquad
    \beta_2\coloneqq\iint_\Theta\!\dd\mu_{\vec{\theta}}\,A(\theta_1)B(\theta_2).
    \label{eq:betas}
\end{equation}
Explicitly, in the circular basis,
\begin{equation}
\hat{\Sigma}_{\varphi_1,\varphi_2|\Theta} =
    \begin{pmatrix}
        1 & \frac{\beta_2}{\bar{X}}\ee^{-2\mathrm{i}\varphi_2} & \frac{\beta_1}{\bar{X}}\ee^{2\mathrm{i}\varphi_1} & \frac{\bar{Y}}{\bar{X}}\ee^{2\mathrm{i}(\varphi_1 - \varphi_2)} \\
        \frac{\beta_2}{\bar{X}}\ee^{2\mathrm{i}\varphi_2} & 1 & \frac{\bar{Y}}{\bar{X}}\ee^{2\mathrm{i}(\varphi_1 + \varphi_2)} & \frac{\beta_1}{\bar{X}}\ee^{2\mathrm{i}\varphi_1} \\
        \frac{\beta_1}{\bar{X}}\ee^{-2\mathrm{i}\varphi_1} & \frac{\bar{Y}}{\bar{X}}\ee^{-2\mathrm{i}(\varphi_1 + \varphi_2)} & 1 & \frac{\beta_2}{\bar{X}}\ee^{-2\mathrm{i}\varphi_2} \\
        \frac{\bar{Y}}{\bar{X}}\ee^{-2\mathrm{i}(\varphi_1 - \varphi_2)} & \frac{\beta_1}{\bar{X}}\ee^{-2\mathrm{i}\varphi_1} & \frac{\beta_2}{\bar{X}}\ee^{2\mathrm{i}\varphi_2} & 1
    \end{pmatrix},
    \label{eq:POVM_Theta_matrix}
\end{equation}
which is \eqnref{eq:Sigma12_matrix} divided by $X$, with $X\to\bar{X}$, $A(\theta_2)B(\theta_1)\to\beta_1$, $A(\theta_1)B(\theta_2)\to\beta_2$ and $Y\to\bar{Y}$. Equivalently, the $\Theta$-integrated cross-section $\iint_\Theta\dd\mu_{\vec{\theta}}\,\frac{\dd^2\sigma(\rho)}{\dd\Omega_1\dd\Omega_2}$ retains the form \eref{eq:gen_cross_section} with the same replacements. At fixed polar angles, $\beta_1\beta_2=XY$, reflecting the product structure $\hat{\Sigma}_{\varphi_1|\theta_1}\otimes\hat{\Sigma}^*_{\varphi_2|\theta_2}$, whereas after averaging in general $\beta_1\beta_2\neq\bar{X}\bar{Y}$, so that $\hat{\Sigma}_{\varphi_1,\varphi_2|\Theta}$ is no longer of product form. Its structure, however, is unchanged, and so are all our results upon the above replacements.

In particular, the $\Delta\varphi$-statistics \eref{eq:av_cross_section} depend on the polar angles only via $\bar{X}$ and $\bar{Y}$, so that $\R$ \eref{eq:R_factor}, $\R_{\max}$ in \thmref{thm:max_r}, $\Rsep$, $\Rsep^{W}$ and $\kappa$ follow upon replacing $X\to\bar{X}$ and $Y\to\bar{Y}$, while the negativity bounds expressed via $\kappa$ remain unchanged. Moreover, as $B(\theta)<0$ for $\theta\in(0,\pi)$, $\beta_1$, $\beta_2$ and $\bar{Y}$ are non-zero, so that the accessible operators are still $\s{\mu}\otimes\s{\nu}$ with $\mu,\nu\in\{0,x,y\}$, and the analysis of \appref{app:witnesses}, including \thmref{thm:R_optimal}, carries over. Only the reconstruction kernels must be adapted: replacing $A(\theta_\ell)/B(\theta_\ell)\to\bar{X}/\beta_\ell$ in $k^{(\ell)}_\mu$ for the local terms, the correlations are reconstructed by
\begin{equation}
    \iint_0^{2\pi}\!\frac{\dd\varphi_1\dd\varphi_2}{(2\pi)^2}\;\frac{\beta_1\beta_2}{\bar{X}\bar{Y}}\,k^{(1)}_i(\varphi_1)\,k^{(2)}_j(\varphi_2)\;\hat{\Sigma}_{\varphi_1,\varphi_2|\Theta}=\s{i}\otimes\s{j},
    \qquad i,j\in\{x,y\},
    \label{eq:kernels_Theta}
\end{equation}
i.e.~with the product $\frac{A(\theta_1)A(\theta_2)}{B(\theta_1)B(\theta_2)}=X/Y$ of the single-photon factors replaced by $\bar{X}/\bar{Y}$, rather than by $\bar{X}^2/(\beta_1\beta_2)$. For instance, the kernel implementing $W_\chi$ involves $\bar{X}/\bar{Y}$ instead of $X/Y$.

Finally, note that 
\begin{equation}
    \frac{\bar{Y}}{\bar{X}}
    =\frac{\iint_\Theta\dd\mu_{\vec{\theta}}\,B(\theta_1)B(\theta_2)}{\bar{X}}
    = \iint_\Theta\dd\mu_{\vec{\theta}} \frac{A(\theta_1)A(\theta_2)}{\bar{X}}\frac{B(\theta_1)B(\theta_2)}{A(\theta_1)A(\theta_2)}
    =\iint_\Theta\!\dd\mu_{\vec{\theta}}\;p(\vec{\theta}|\Theta)\,\frac{Y(\vec{\theta})}{X(\vec{\theta})}
\end{equation}
is the average of $Y/X$ over the polar angles of the selected events, rather than $Y/X$ evaluated at the centre of $\Theta$, so that $\bar{Y}/\bar{X}\le\max_\Theta Y/X$, i.e.~averaging cannot yield a larger azimuthal modulation than the largest one within $\Theta$. 

As $Y/X$ is maximal at $\theta_1=\theta_2\approx82^\circ$, averaging over a disc centred there reduces the modulation:~for radius $r=20^\circ$, $\bar{Y}/\bar{X}\approx0.430$ compared to $Y/X\approx0.479$ at the centre, which lowers $\Rsep$ from $\approx1.629$ to $\approx1.547$, as shown in the inset of \figref{fig:analytical_pred}. Moreover, as $\R$ is fixed by the measurement, replacing $\bar{Y}/\bar{X}$ by $\max_\Theta Y/X$ yields a lower bound on the value of $\kappa=\frac{\bar{X}}{\bar{Y}}\frac{\R-1}{\R+1}$ inferred from $\R$, and hence $\Rsep$-thresholds and negativity bounds valid irrespective of the weighting of polar angles within $\Theta$, e.g.~due to varying detection efficiency $\eta(\vec{\theta})$.

\end{document}